\documentclass[aps,prd,reprint,amsmath,amssymb,nofootinbib,longbibliography]{revtex4-2}
\usepackage[T1]{fontenc}
\usepackage{lmodern}
\usepackage[english]{babel}
\addto\captionsenglish{}
\usepackage{graphicx,xcolor}
\usepackage{amsthm}
\usepackage{microtype}
\usepackage[colorlinks=true,linkcolor=blue!45!black,citecolor=blue!45!black,urlcolor=blue!45!black]{hyperref}
\hypersetup{pdftitle={CPL dark energy: exact pressure–density relation and limits on cosmic acceleration},pdfauthor={Ozgur Akarsu, Maria Caruana, Luis A. Escamilla, Emre O. Kahya}}
\usepackage{orcidlink}
\newcommand{\dd}{\mathrm d}
\newcommand{\rde}{\rho_{\rm de}}
\newcommand{\pde}{p_{\rm de}}
\newcommand{\Ide}{\mathcal I_{\rm de}}
\newcommand{\Mde}{\mathcal M_{\rm de}}
\newcommand{\Ibg}{\mathcal I_{\rm bg}}
\newcommand{\Mbg}{\mathcal M_{\rm bg}}
\newcommand{\Itot}{\mathcal I_{\rm tot}}
\newcommand{\Mtot}{\mathcal M_{\rm tot}}
\newcommand{\Wi}{W_{-1}}
\newcommand{\Uc}{\mathcal U}
\newcommand{\Om}{\Omega_{m0}}
\newcommand{\Ode}{\Omega_{{\rm de}0}}
\newcommand{\Kmax}{K_{\max}}
\newcommand{\Jmax}{J_{\max}}
\newcommand{\phihat}{\hat\varphi}
\newcommand{\psihat}{\hat\psi}

\newtheorem{theorem}{Theorem}
\newtheorem{lemma}{Lemma}
\newtheorem{corollary}{Corollary}

\makeatletter
\def\bibsection{\par\begingroup\baselineskip26\p@\bib@device{\hsize}{72\p@}\endgroup\nobreak\@nobreaktrue\addvspace{19\p@}}
\makeatother
\begin{document}

\title{CPL dark energy: exact pressure--density relation and limits on cosmic acceleration}
\author{\"{O}zg\"{u}r Akarsu\,\orcidlink{0000-0001-6917-6176}}
\email{akarsuo@itu.edu.tr}
\affiliation{Department of Physics, Istanbul Technical University, Maslak 34469 Istanbul, T\"{u}rkiye}
\author{Maria Caruana\,\orcidlink{0000-0002-8989-0462}}
\email{caruanamaria@itu.edu.tr}
\affiliation{Department of Physics, Istanbul Technical University, Maslak 34469 Istanbul, T\"{u}rkiye}
\author{Luis A. Escamilla\,\orcidlink{0000-0003-4334-5140}}
\email{torresl@itu.edu.tr}
\affiliation{Department of Physics, Istanbul Technical University, Maslak 34469 Istanbul, T\"{u}rkiye}
\author{Emre O. Kahya\,\orcidlink{0000-0003-2760-7091}}
\email{eokahya@itu.edu.tr}
\affiliation{Department of Physics, Istanbul Technical University, Maslak 34469 Istanbul, T\"{u}rkiye}
%\date{}

\begin{abstract}
The Chevallier--Polarski--Linder (CPL) parametrization, $w=w_0+w_a(1-a)$, is widely used to assess current indications of evolving dark energy. We ask what the parametrization itself imposes, independently of the data. First, we invert its pressure--density relation in closed form using the real branches of the Lambert $W$ function, and classify its positive-density histories with $w_a\ne0$ into six classes. Every history that crosses $w=-1$ reduces, in normalized variables, to one parameter-free curve, folded at a density extremum: the equation of state has two parameters, and the third only locates the present on the curve. No single-valued barotropic fluid follows the curve through the fold, so a crossing requires a continuation prescription. Second, we prove that a separately conserved, positive-density component accompanied by dust, radiation and collisionless massive species produces at most one accelerating epoch on a connected expanding branch, for any spatial curvature, if $w$ is nondecreasing and convex in $\ln a$, or nonincreasing; counterexamples delimit these hypotheses. For CPL with $w_a<0$, the case relevant to the DESI mean values, we give the exact dust-limit condition for an accelerating window to exist and an accurate approximation to its edges. At the DESI DR2 mean values the window is still open today for the three combinations that include supernovae, and closed at $z\simeq0.07$ for BAO+CMB alone. Conditional on $w_a<0$ and continued CPL evolution, any accelerating window is bounded; observations constrain its existence and location.

\end{abstract}
\maketitle

\section{Introduction}
The Chevallier--Polarski--Linder (CPL) parametrization \cite{Chevallier:2000qy,Linder:2002et}
\begin{equation}
 w(a)=w_0+w_a(1-a)=w_i-w_a a,\qquad w_i\equiv w_0+w_a,
 \label{eq:cpl}
\end{equation}
is widely used to test the idea of a time-evolving dark energy. Within CPL, the original DESI DR2 analysis of baryon acoustic oscillation (BAO) measurements combined with the cosmic microwave background (CMB) reports a preference for dynamical dark energy over a cosmological constant at $3.1\sigma$, and at $2.8$--$4.2\sigma$ for the combinations with supernova (SN) samples \cite{DESI:2025zgx}. A subsequent DES-Dovekie reanalysis, combined with DESI DR2 BAO and updated CMB likelihoods, reports a frequentist preference of $3.2\sigma$ for CPL over $\Lambda$CDM \cite{DES:2025sig}. The preferred region has $w_0>-1$, $w_a<0$ and $w_i<-1$, and therefore a crossing of the line $w=-1$ \cite{Ozulker:2025ehg}. Present-day acceleration is nevertheless not guaranteed: the BAO+CMB inference is compatible with $q_0\ge0$ \cite{Colgain:2025nzf,Wang:2025owe,Xu:2026sbw,Ferri:2026sat}. Related reports of slowing or transient acceleration predate DESI \cite{Shafieloo:2009ti,Cardenas:2013roa,Magana:2017usz} and occur in other dark-energy models \cite{deSouza:2025rhv,Ferrario2026}.

The CPL-based analyses quoted above motivate two questions about what the parametrization itself imposes, independently of the data: which pressure--density histories does CPL encode, and which acceleration histories does it permit? Whenever $w$ crosses $-1$, CPL encodes a folded, two-branch equation of state with two parameters, given in closed form by Eq.~\eqref{eq:eos}. Under the source assumptions of Corollary~\ref{cor:cpl}, CPL with $w_a<0$ permits at most one accelerating epoch on a connected expanding branch, confined to a bounded interval in scale factor. Observations constrain the parameters that determine whether such a window exists and where the present lies relative to it.

On the first question, Scherrer \cite{Scherrer:2015tra} obtained the implicit equation of state and plotted pressure against density over ranges with $w\ge-1$; a folded CPL trajectory appears in a later plot by Sadri \cite{Sadri:2017ezh}. Our contribution is the explicit inversion with its real domains, the chronology of its two branches, a classification of the positive-density histories with $w_a\ne0$ into six classes, and a normalized representation that separates the equation-of-state parameters from the position along a history. A background pressure--density relation need not define a regular autonomous fluid. This distinction connects our branch-continuation analysis to kinetic k-essence reconstructions and barotropic-fluid studies of the physical admissibility of such mappings \cite{dePutter:2007ny,Cardenas:2018ltl,Perkovic:2020mph,Perkovic:2020eju,Akarsu:2015wqa}. At the fold it becomes concrete: an incoming history may wait there for any number of e-folds before leaving along the second branch; leaving at once is uninterrupted CPL, and never leaving reproduces, for a past crossing, a restricted model of Ref.~\cite{Ozulker:2025ehg}.

The second question concerns the distinction between properties imposed by a parametrization and information supplied by the data \cite{Nesseris:2025lke,Wolf:2023uno,Shlivko:2024llw}. For example, marginalizing over quadratic and cubic terms in $1-a$ weakens the evidence for phantom behavior in the DR1-based analysis of Ref.~\cite{Nesseris:2025lke}. We address the number and structure of acceleration epochs using the two dark-energy source combinations
\begin{equation}
 \Ide\equiv\rde+\pde,\qquad \Mde\equiv\rde+3\pde,
 \label{eq:diagnostics}
\end{equation}
the inertial and active gravitational mass densities. In the defocusing framework of Ref.~\cite{Akarsu:2026pia}, $\Ide<0$ identifies violation of the null energy condition (NEC) by the dark-energy sector, while $\Mde<0$ identifies a repulsive contribution to the comoving Raychaudhuri equation and, for $\rde>0$, violation of the strong energy condition (SEC) by that sector \cite{Raychaudhuri:1953yv,Hawking:1973uf}. Cosmic acceleration is determined by the \emph{total} active density, $\Mde+\Mbg$, where $\Mbg>0$ is that of the accompanying dust, radiation and collisionless massive species; it is negative where $K\equiv-\Mde/\Mbg>1$. We prove that, where it is positive, $K$ is log-concave if $w$ is nondecreasing and convex in $\ln a$, and increasing if $w$ is nonincreasing. The total is then negative on at most one interval, in particular for CPL; counterexamples arise when the conditions on $w$ or on that content are relaxed.

Sections~\ref{sec:eos} and \ref{sec:fluid} develop the inverse equation of state, the six classes and their fluid interpretation. Section~\ref{sec:theorem} proves the acceleration theorem; Secs.~\ref{sec:cpl} and \ref{sec:plane} derive the CPL thresholds and evaluate them at the DESI DR2 mean values. The inversion and the theorem are logically independent, but meet through $\Ide$, $\Mde$ and the pressure identities~\eqref{eq:pressureid}: an accelerating window is an arc of the pressure--density trajectory [Fig.~\ref{fig:eos}(a)]. Appendices~\ref{app:ratios} and \ref{app:asymptotic} extend the inversion to density ratios and give the asymptotic forms of the equation of state.

\emph{Scope and conventions.} We set $c=1$, normalize $a=1$ today, and use primes for $\dd/\dd N$, where $N\equiv\ln a$. The equation-of-state analysis assumes a homogeneous, separately conserved component with $\rde>0$ on a Friedmann--Lema\^itre--Robertson--Walker (FLRW) background; general relativity enters in Sec.~\ref{sec:theorem}. The density normalization is included in the parameter count, $(w_0,w_a,\rho_{{\rm de}0})$. A regular single-valued barotrope with finite adiabatic sound speed, or a single canonical minimally coupled scalar, cannot realize a phantom crossing; effective fluids require an additional perturbative closure or field realization \cite{Vikman:2004dc,Hu:2004kh,Kunz:2006wc}. Three qualifications apply throughout and are stated here once. (i)~CPL is a truncated expansion about $a=1$: statements about early times and future epochs describe the ansatz continued beyond the range in which it is fitted. (ii)~The sets of scale factors on which the total active or inertial density is negative are properties of the sources alone; an expanding solution realizes the part of them that it reaches, which is all of it in flat and open models whose densities are all positive. (iii)~The DESI DR2 numbers are the published mean values (Table~\ref{tab:desi}); they illustrate the analytic results and are not posterior constraints on the derived quantities. The \emph{reference history} is the BAO+CMB mean, $(w_0,w_a,\Om)=(-0.42,-1.75,0.353)$ \cite{DESI:2025zgx}, with a flat dust background unless stated otherwise.

\section{The equation of state of CPL in closed form}
\label{sec:eos}
\subsection{Explicit solution}
\label{sec:explicit}
Conservation, $\rde'=-3(\rde+\pde)$, gives
\begin{equation}
 \varrho\equiv\frac{\rde}{\rho_{{\rm de}0}}=a^{-3s}e^{3w_a(a-1)},\qquad s\equiv1+w_i ,
 \label{eq:background}
\end{equation}
with $\pde=w\rde$. The equation of state $\pde(\rde)$ is obtained by eliminating $a$ between Eqs.~\eqref{eq:cpl} and \eqref{eq:background}. Let $w_a\ne0$ and $s\ne0$, and define the signed number
\begin{equation}
 a_I\equiv\frac{s}{w_a}=\frac{1+w_0+w_a}{w_a},
 \label{eq:aI}
\end{equation}
so that $1+w=s\,(1-a/a_I)$; when $a_I>0$ it is the scale factor at which $w=-1$. Raising Eq.~\eqref{eq:background} to the power $-1/(3s)$ gives $a\,e^{-a/a_I}=e^{-1/a_I}\varrho^{-1/(3s)}$, that is,
\begin{equation}
 \Bigl(-\frac{a}{a_I}\Bigr)e^{-a/a_I}=Z,\qquad Z\equiv-\frac{e^{-1/a_I}}{a_I}\,\varrho^{-1/(3s)} .
 \label{eq:Z}
\end{equation}
This is the defining relation $We^{W}=Z$ of the Lambert function \cite{Corless:1996zz}. Hence $a=-a_I\,W_k(Z)$ and, since $w=w_i-w_aa$,
\begin{equation}
 \boxed{\begin{gathered}
 \pde=\rde\,\bigl[\,w_i+s\,W_k(Z)\bigr],\\
 Z=-\frac{w_a}{s}\,e^{-w_a/s}\Bigl(\frac{\rde}{\rho_{{\rm de}0}}\Bigr)^{-1/(3s)},
 \end{gathered}}
 \label{eq:eos}
\end{equation}
where $w_i=w_0+w_a$ and $s=1+w_0+w_a$. Equivalently $1+w=s\,[1+W_k(Z)]$, and $a=-(s/w_a)\,W_k(Z)$ is the scale factor at which the dark energy has a given density. Both real branches of $W$ are available in standard numerical libraries, so Eq.~\eqref{eq:eos} can be evaluated directly.

\begin{figure*}[t]
 \centering
 \includegraphics[width=\textwidth]{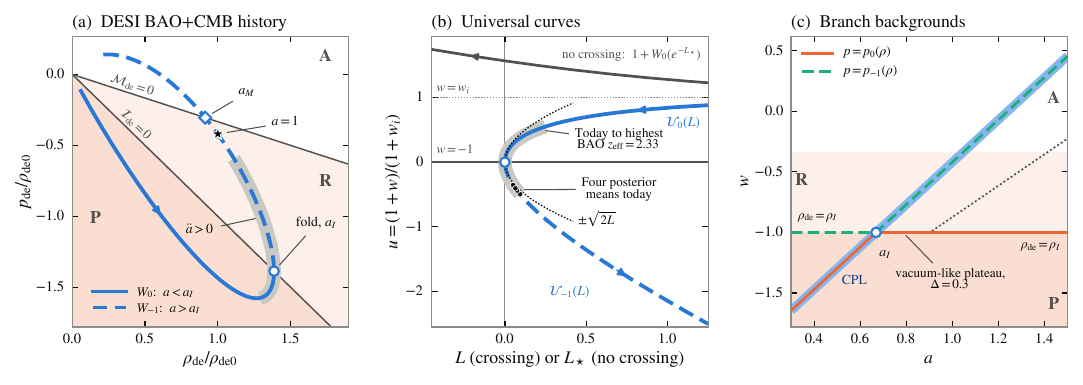}
 \caption{The equation of state of CPL. (a)~Pressure--density trajectory of the reference history, the DESI DR2 BAO+CMB mean \cite{DESI:2025zgx}, for $0.12\le a\le1.65$, in units of the present density. The solid and dashed curves are the branches $k=0$ and $k=-1$ of Eq.~\eqref{eq:eos}, joined at the fold (circle), where $w=-1$; the diamond and the star mark $a_M$ and $a=1$, and arrows, here and in (b), point toward increasing $a$. Shading marks the sectors of Eq.~\eqref{eq:sectors}, bounded by $\Mde=0$ and $\Ide=0$. The gray band is the arc on which the expansion accelerates [Eq.~\eqref{eq:accwindow}, Sec.~\ref{sec:plane}]. (b)~The universal crossing curve~\eqref{eq:universal}, plotted against $L$, and its leading behavior $\pm\sqrt{2L}$ (dotted). The gray band marks the $L$ values reached between today and the highest BAO effective redshift, $z_{\rm eff}=2.33$, for the reference history. Black marks locate the present for the four DESI DR2 mean values (Table~\ref{tab:desi}). The upper curve represents histories without a crossing, Eq.~\eqref{eq:nocross}; its horizontal coordinate is $L_\star$, with the distinct density normalization defined there. (c)~The two single-branch barotropic backgrounds of Sec.~\ref{sec:branches} for the reference history: each follows CPL (thick line) on one side of $a_I$ and has a constant dark-energy density on the other. The dotted solution waits for $\Delta=0.3$ e-folds at the vacuum-like state before departing along the second branch.}
 \label{fig:eos}
\end{figure*}

The index $k$ labels the real branches of $W$: the principal branch $W_0\ge-1$, defined for $Z\ge-1/e$, and $\Wi\le-1$, defined for $-1/e\le Z<0$. They meet at $Z=-1/e$, where $W=-1$. Which of them occurs is decided by the sign of $a_I$.

(a) If $a_I<0$, $w$ never equals $-1$. Then $Z>0$, only $W_0$ is real, the density is monotonic and $\pde(\rde)$ is single valued, with $k=0$.

(b) If $a_I>0$, then $-1/e\le Z<0$ and both branches are needed, $k=0$ for $a\le a_I$ and $k=-1$ for $a\ge a_I$. They join at $a_I$, where $Z=-1/e$, $w=-1$ and the density has its extremum,
\begin{equation}
 \rho_I\equiv\rde(a_I)=\rho_{{\rm de}0}\,a_I^{-3s}\,e^{3s(1-1/a_I)} .
 \label{eq:rhoI}
\end{equation}
For $s>0$ this is a minimum, with $\rde\ge\rho_I$; for $s<0$ it is a maximum, with $0<\rde\le\rho_I$. Every other attained density occurs once on each branch.

(c) The excluded cases are elementary. For $w_a=0$, $\pde=w_0\rde$. For $s=0$ and $w_a\ne0$, $\varrho=e^{3w_a(a-1)}$ and
\begin{equation}
 \pde=\rde\,\bigl(w_0-\tfrac13\ln\varrho\bigr),
 \label{eq:log}
\end{equation}
which is single valued and is the limit of Eq.~\eqref{eq:eos} as $s\to0$ on the branch corresponding to the physical history. Its density domain is fixed by $a=1+\ln\varrho/(3w_a)>0$.

Table~\ref{tab:sectors} lists the six classes of history to which these cases correspond (Sec.~\ref{sec:classes}). Equation~\eqref{eq:eos} can be verified directly by substitution, $W_k(Z)=-a/a_I$.   Nothing in it is tied to the exponent of $a$ in Eq.~\eqref{eq:cpl}: for $w=w_i-w_na^{n}$ with any $n\ne0$, and $s=1+w_i$ as before, Eq.~\eqref{eq:eos} holds with $w_a\to w_n$ and $\varrho^{-1/(3s)}\to\varrho^{-n/(3s)}$ in $Z$; the case $n=-1$ is an equation of state linear in redshift. Nor is the inversion special to the density itself: the same steps apply to the ratio of $\rde$ to any density with a constant equation of state, and with dust they give the epochs of matter--dark-energy equality in closed form (Appendix~\ref{app:ratios}).

\emph{Relation to earlier work.} Scherrer's Eqs.~(25) and (29) give the implicit relation, and his Figs.~4--5 plot CPL pressure--density curves over $0.3\le a\le1$ \cite{Scherrer:2015tra}; his parameter choices keep $w\ge-1$ there, so those curves are single valued. He showed that, in units of the characteristic density $\rho_C\equiv\rho_{{\rm de}0}|w_a|^{3s}e^{-3w_a}$, the implicit locus depends on $w_i$ alone; its physical component additionally obeys ${\rm sgn}(w_i-w)={\rm sgn}(w_a)$, since $w_i-w=w_aa$ and $a>0$. A folded curve appears, without comment, in Fig.~5 of Ref.~\cite{Sadri:2017ezh}, drawn for $(w_0,w_a)=(-1.09,0.43)$. The curves and their multivaluedness therefore have precedents. Equation~\eqref{eq:eos} resolves their real branches and domains explicitly. Kinetic reconstructions instead display $F(X)$ \cite{dePutter:2007ny,Cardenas:2018ltl}, which, as Ref.~\cite{dePutter:2007ny} notes, becomes double valued where $w$ crosses $-1$ or the pressure is extremal; fluid admissibility has been studied through $\dd p/\dd\rho$ \cite{Perkovic:2020mph,Perkovic:2020eju}. The obstruction to regular barotropic crossing is general \cite{Stefancic:2005cs,Kunz:2006wc}, and analogous real Lambert branches occur in logotropic fluids \cite{Ferreira:2016goc}. We are not aware of an earlier explicit CPL inversion of the form~\eqref{eq:eos}.

\begin{table}[t]
\caption{The six classes of CPL history for $\rde>0$ and $w_a\ne0$, for increasing $a\in(0,\infty)$. Sectors A, R, P are defined in Eq.~\eqref{eq:sectors} and the crossing epochs in Eqs.~\eqref{eq:aI} and \eqref{eq:crossings}. Classes with a fold have the two-branch equation of state and lie on the universal curve~\eqref{eq:universal}; the others are single valued and obey Eq.~\eqref{eq:nocross} for $s\ne0$. The $s=0$ boundaries in classes III and V instead obey Eq.~\eqref{eq:log}. All four DESI DR2 mean values (Table~\ref{tab:desi}) belong to class~IV.}
\label{tab:sectors}
\begin{ruledtabular}
\begin{tabular}{clccc}
Class & Parameters & Sectors & Crossings & Fold\\
\hline
I & $w_a>0,\ w_i>-\tfrac13$ & A$\to$R$\to$P & $a_M<a_I$ & min.\\
II & $w_a>0,\ -1<w_i\le-\tfrac13$ & R$\to$P & $a_I$ & min.\\
III & $w_a>0,\ w_i\le-1$ & P & none & ---\\
IV & $w_a<0,\ w_i<-1$ & P$\to$R$\to$A & $a_I<a_M$ & max.\\
V & $w_a<0,\ -1\le w_i<-\tfrac13$ & R$\to$A & $a_M$ & ---\\
VI & $w_a<0,\ w_i\ge-\tfrac13$ & A & none & ---
\end{tabular}
\end{ruledtabular}
\end{table}

\subsection{One universal curve}
\label{sec:universal}
Measure the density from its extremum and the equation of state in units of $s$,
\begin{equation}
 L\equiv\frac{1}{3s}\ln\frac{\rde}{\rho_I},\qquad u\equiv\frac{1+w}{s}=1-x,\qquad x\equiv\frac{a}{a_I}.
 \label{eq:Lu}
\end{equation}
For $a_I>0$, Eqs.~\eqref{eq:background} and \eqref{eq:rhoI} give $\rde/\rho_I=(x\,e^{1-x})^{-3s}$, so that
\begin{equation}
 L=x-1-\ln x\;\ge0
 \label{eq:normal}
\end{equation}
whatever the sign of $s$, and $Z=-e^{-1-L}$. Equation~\eqref{eq:eos} becomes
\begin{equation}
 \boxed{\;u=\Uc_k(L)\equiv1+W_k\bigl(-e^{-1-L}\bigr),\qquad k=0,-1,\;}
 \label{eq:universal}
\end{equation}
in which no parameter appears [Fig.~\ref{fig:eos}(b)]. Every CPL history with a crossing traces this curve in the same way. It descends the principal branch $\Uc_0$ from $u\to1$ ($w\to w_i$) at $L\to\infty$ to the fold, $u=0$ at $L=0$, and leaves along $\Uc_{-1}$, on which $u\to-\infty$. In the plane $(\tfrac13\ln(\rde/\rho_{\rm ref}),\,1+w)$, with a fixed positive reference density $\rho_{\rm ref}$, the curves are related by a translation and a dilation by $s$ about the fold. The dilation includes a point reflection when $s<0$; the fold is a minimum of the density for $s>0$ and a maximum for $s<0$. The same curve serves the family $w=w_i-w_na^n$, with $L\to nL$ and the direction of traversal reversed for $n<0$.

Equation~\eqref{eq:universal} displays the parameter content of CPL. As an equation of state,
\begin{equation}
 \pde=\rde\Bigl[-1+s\,\Uc_k\Bigl(\frac{1}{3s}\ln\frac{\rde}{\rho_I}\Bigr)\Bigr],
 \label{eq:twoparam}
\end{equation}
it has two parameters, $s$ and $\rho_I$. The full background specification $(\rho_{{\rm de}0},w_0,w_a)$ is thus equivalently $(\rho_I,s,a_I)$ for histories with a crossing. Once $s$ and $\rho_I$ are fixed, $a_I$ is the integration constant of the continuity equation that locates the present on the curve:
\begin{equation}
 u_0=1-\frac1{a_I},\qquad L_0=\frac1{a_I}-1+\ln a_I .
 \label{eq:today}
\end{equation}
The pair $(w_0,w_a)=(s-1-s/a_I,\,s/a_I)$ mixes the two kinds of parameter. In the $(w_0,w_a)$ plane, $a_I$ is constant on the lines $1+w_0=(a_I-1)\,w_a$ through the cosmological-constant point, the crossing-epoch geometry discussed in Ref.~\cite{Ozulker:2025ehg}. Each line consists of two rays distinguished by the sign of $s$. Models at fixed $a_I$ occupy the same point of the universal curve today; after measuring density from $\rho_I$, their curves differ by the dilation $s$. At fixed $a_I$, the limit $s\to0$ gives a cosmological constant. For crossing histories, after centering at the fold, the $w_i$ dependence that remains in Scherrer's scaled relation \cite{Scherrer:2015tra} is therefore this dilation alone.

For histories without a crossing ($a_I<0$) the same steps, with $\rho_\star\equiv\rho_{{\rm de}0}|a_I|^{-3s}e^{3s/|a_I|}$, the density at $a=W_0(1)\,|a_I|\simeq0.567\,|a_I|$, and $L_\star\equiv\ln(\rde/\rho_\star)/(3s)$, give
\begin{equation}
 u=1+W_0\bigl(e^{-L_\star}\bigr),\qquad L_\star\in\mathbb R,
 \label{eq:nocross}
\end{equation}
again without parameters: $u$ grows monotonically from $1$ as $L_\star$ falls from $+\infty$ to $-\infty$.

\emph{The fold.} Near $L=0$,
\begin{equation}
 \Uc_k(L)=\pm\sqrt{2L}\,\Bigl(1+\frac{L}{18}\Bigr)-\frac{2L}{3}+O(L^2),
 \label{eq:foldexp}
\end{equation}
with the upper sign for $k=0$. The leading term is the normal form of a fold and is not special to CPL. For any smooth $w(N)$ with a simple zero of $1+w$ at $N_I$, with $\delta N\equiv N-N_I$ and $\rho_I\equiv\rde(N_I)$,
\begin{equation}
 \begin{aligned}
 \frac{\rde-\rho_I}{\rho_I}&=-\frac32w'_I\,\delta N^2+O(\delta N^3),\\
 \frac{\pde+\rho_I}{\rho_I}&=w'_I\,\delta N+O(\delta N^2),
 \end{aligned}
 \label{eq:foldlocal}
\end{equation}
so that the trajectory is smooth, has a vertical tangent in the $(\rde,\pde)$ plane, and $\pde+\rho_I\propto\pm|\rde-\rho_I|^{1/2}$. What CPL supplies is the whole curve. Its two branches are tied together in a simple way: a density that occurs at $a_1<a_I$ occurs again at $a_2>a_I$, with
\begin{equation}
 \Uc_0(L)-\Uc_{-1}(L)=\ln\frac{a_2}{a_1}=\frac{a_2-a_1}{a_I}.
 \label{eq:logmean}
\end{equation}
The gap between the two values of the equation of state at a given density, in units of $s$, is the number of e-folds that separates the two epochs, and the fold epoch is their logarithmic mean.

\emph{DESI DR2 mean values.} The four mean values of Table~\ref{tab:desi} lie on the outgoing branch at $L_0=0.05$--$0.09$. Between today and the highest BAO effective redshift, $z_{\rm eff}=2.33$, their histories remain at $L<0.31$ [Fig.~\ref{fig:eos}(b)]. There Eq.~\eqref{eq:foldexp} has relative error below $0.25\%$, whereas its leading square root differs by about $30\%$ at the high-redshift endpoint. These are errors in the normalized equation of state; distances require the background integrals. Crossing-epoch and density-extremum descriptions have also been studied in Refs.~\cite{Ozulker:2025ehg,DiValentino:2020naf}.

\subsection{Sectors and six classes}
\label{sec:classes}
For $\rde>0$ the diagnostics~\eqref{eq:diagnostics} define three sign sectors,
\begin{equation}
 \begin{array}{lll}
 {\rm A}:&\Ide>0,\ \Mde>0&(w>-\tfrac13),\\[1pt]
 {\rm R}:&\Ide>0,\ \Mde<0&(-1<w<-\tfrac13),\\[1pt]
 {\rm P}:&\Ide<0,\ \Mde<0&(w<-1),
 \end{array}
 \label{eq:sectors}
\end{equation}
attractive, repulsive with the NEC satisfied, and NEC violating (phantom-like); the fourth combination is excluded by $\Mde=3\Ide-2\rde$. For CPL the normalized diagnostics $\Ide/\rde$ and $\Mde/\rde$ are linear in $a$,
\begin{gather}
 \Ide=w_a(a_I-a)\rde,\qquad \Mde=3w_a(a_M-a)\rde,\label{eq:linear}\\
 a_M\equiv\frac{1+3w_i}{3w_a}=a_I-\frac{2}{3w_a}.
 \label{eq:crossings}
\end{gather}
We treat $a_I$ and $a_M$ as signed numbers: the NEC (SEC) boundary is crossed if and only if $a_I>0$ ($a_M>0$). Each factor in Eq.~\eqref{eq:linear} changes sign at most once, so Table~\ref{tab:sectors} lists every history. Classes I, II and IV have a fold; on the universal curve the NEC boundary is the fold itself and the SEC boundary is the point $u=2/(3s)$. When both crossings occur the SEC boundary is crossed first for $w_a>0$ and last for $w_a<0$, and $\Mde(a_I)=-2\rde<0<\tfrac23\rde=\Ide(a_M)$: the NEC crossing lies inside the repulsive region.

\begin{figure*}[t]
 \centering
 \includegraphics[width=\textwidth]{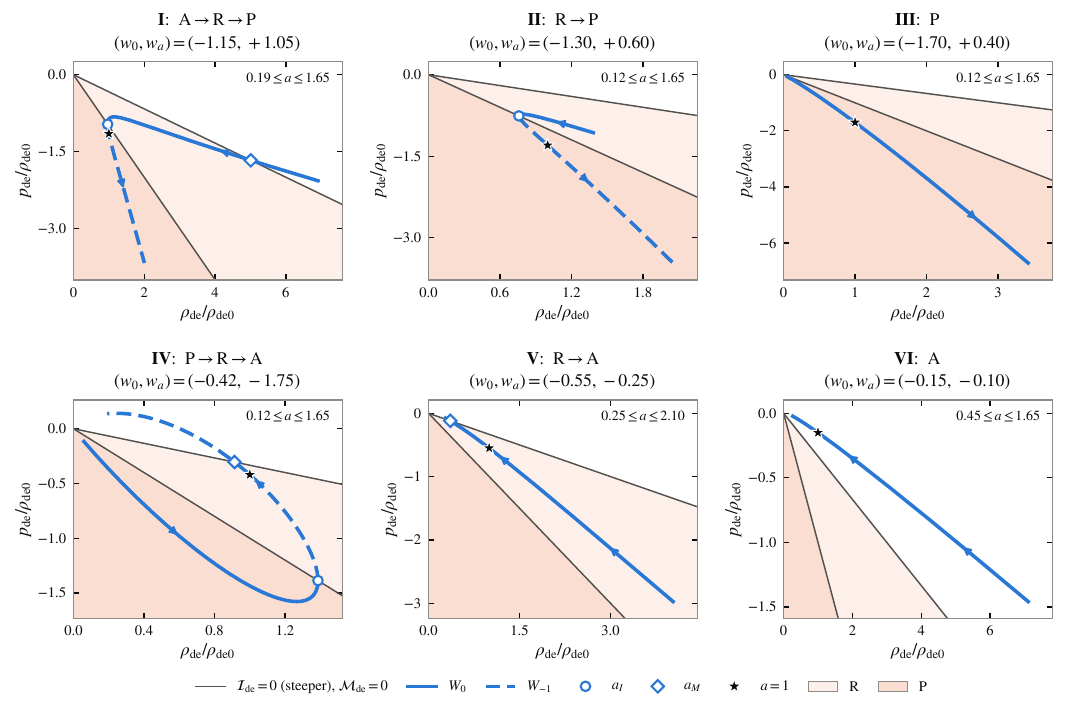}
 \caption{Representative pressure--density trajectories for all six classes of Table~\ref{tab:sectors}. Parameters and plotted scale-factor ranges appear in each panel; arrows point toward increasing $a$. Both axes use the fixed present dark-energy density of that history. Solid and dashed curves identify $W_0$ and $W_{-1}$ in Eq.~\eqref{eq:eos}. Circles mark the density folds at $\Ide=0$: minima in I and II, a maximum in IV. Diamonds mark $\Mde=0$, and stars mark today. The steep and shallow lines are $p=-\rho$ and $p=-\rho/3$; shading distinguishes sectors P and R, with A above the latter line. These are boundaries for the component alone, not the acceleration boundary of a universe containing other matter. Panel~IV is the reference history; the other parameter values are chosen for illustration.}
 \label{fig:atlas}
\end{figure*}

Figure~\ref{fig:atlas} displays one representative of each class. These are six classes of history, not six Lambert branches: I and II have a density minimum, IV a maximum, and III, V and VI are monotonic. All folds are traversed from $W_0$ to $W_{-1}$ as $a$ increases. The atlas shows both directions of phantom crossing and the order of the inertial- and active-density sign changes; the four DESI DR2 mean values all lie in class~IV.

Equations~\eqref{eq:aI} and \eqref{eq:crossings} can be inverted: $w_a=2/[3(a_I-a_M)]$ and $s=w_aa_I$. The two crossing epochs therefore parametrize CPL,
\begin{equation}
 1+w(a)=\frac23\,\frac{a_I-a}{a_I-a_M}.
 \label{eq:reparam}
\end{equation}
In class~IV, $a_I$ is the phantom-crossing epoch and $a_M$ the epoch at which the component stops being repulsive. When both crossings occur, the two epochs and the two densities obey a relation free of parameters,
\begin{equation}
 \ln\frac{\rde(a_M)}{\rde(a_I)}=-2-\frac{2\ln\chi}{1-\chi},\qquad\chi\equiv\frac{a_M}{a_I}.
 \label{eq:consistency}
\end{equation}
For the reference history [Fig.~\ref{fig:eos}(a)], $a_I=0.669$ and $a_M=1.050$: the density peaked at $1.387\,\rho_{{\rm de}0}$ at $z_I=0.496$ and falls to $0.659$ of that value by the time the component stops being repulsive.

Finally, reversing the sign of $\rho_{{\rm de}0}$ reflects every trajectory through the origin and reverses the sector signs. Such a history cannot be reached from $\rde>0$: a conserved component changes the sign of its density only where $w$ diverges \cite{Ozulker:2022slu}, so the sign-switching histories of Refs.~\cite{Akarsu:2019hmw,Akarsu:2021fol,Akarsu:2022typ,Akarsu:2026pia}, and the sign-switching extensions of CPL of Ref.~\cite{Gokcen:2026pkq}, are not continuations of a CPL curve.

\section{CPL as a fluid}
\label{sec:fluid}
The inverse relation distinguishes a prescribed CPL history from an autonomous fluid closure. This section treats each branch as a fluid in its own right and shows why a crossing needs a prescription. Appendix~\ref{app:asymptotic} gives the asymptotic forms of Eq.~\eqref{eq:eos} before and after the crossing: a constant-$w$ fluid with polytropic corrections at early times and a logarithmic fluid at late times.

\subsection{Autonomous branch backgrounds}
\label{sec:branches}
Consider a barotropic background with one branch of Eq.~\eqref{eq:twoparam} as its equation of state, $p=p_k(\rho)$ with $k$ fixed; whether it defines a regular physical fluid is a separate, perturbative question (Sec.~\ref{sec:fold}). Its continuity equation is the autonomous flow
\begin{equation}
 \frac{\dd L}{\dd N}=-\,\Uc_k(L).
 \label{eq:flow}
\end{equation}
The fold $L=0$ is a vacuum-like equilibrium of this component, with $p=-\rho=-\rho_I$. This does not by itself imply a de Sitter spacetime in the presence of other matter. Because $\Uc_k\simeq\pm\sqrt{2L}$, the flow is not Lipschitz continuous there. A nonconstant branch reaches or leaves the equilibrium after a finite interval in $N$: by Eq.~\eqref{eq:Lu},
\begin{equation}
 |N-N_I|=\bigl|\ln\bigl[1-\Uc_k(L)\bigr]\bigr| ,
 \label{eq:efolds}
\end{equation}
where $N_I$ is the arrival or departure epoch of the nonconstant branch; the interval is $\simeq\sqrt{2L}$ near the fold.
The branches behave differently [Fig.~\ref{fig:eos}(c)]. For $k=0$, $L$ decreases: the background follows CPL up to $a_I$ and then stays at $L=0$, since $L\ge0$ and $L'\le0$. For $k=-1$, it may remain at $L=0$ for an arbitrary number of e-folds before departing along a CPL segment. The forward solution through the equilibrium is therefore unique on $W_0$ but not on $W_{-1}$. The equilibrium attracts the former branch in a finite e-fold interval and repels the latter; neither fixed-branch background crosses $w=-1$.

For $0<a_I<1$, these backgrounds correspond, up to density normalization, to CPL$_{<a_c}$ and CPL$_{>a_c}$ of \"Oz\"ulker, Di~Valentino and Giar\`e \cite{Ozulker:2025ehg}, with $a_c=a_I$: CPL on one side and constant density on the other. Outside that past-crossing interval, their definitions revert to unmodified CPL. Here these restricted histories arise as the solutions of the single-branch equations of state, the departure epoch on $W_{-1}$ being an integration constant.

CPL itself changes branch on arrival at the fold. Fix a nonconstant incoming $W_0$ history, including its integration constant $a_I$, and require a continuous pressure. Its continuations on the two-branch relation consist of a vacuum-like plateau followed by $W_{-1}$, or a plateau of infinite duration. For a finite plateau, its duration is $\Delta\equiv\ln(a_{\rm out}/a_I)\ge0$ e-folds, where $a_{\rm out}$ is the departure scale factor.   The outgoing segment again has the CPL form, with the same $w_i=w_0+w_a$ and $w_a\to w_ae^{-\Delta}$; its density normalization is fixed by continuity at $\rho_I$. Among continuations that eventually leave the fold, uninterrupted CPL, $\Delta=0$, is the unique one with differentiable $w(a)$. For $\Delta>0$ the plateau has $w'=0$, whereas each adjoining CPL segment has $w'=-s\ne0$ at the join. The pure equilibrium $w=-1$, $\rho=\rho_I$ is another smooth solution of the relation, but does not match the specified nonconstant incoming history.   The restricted histories are obtained by extending the plateau indefinitely into the future, or into the past with a fixed outgoing segment. Crossing thus requires a branch-selection prescription; it is not determined by a single-valued barotropic equation of state.

In Table~II of Ref.~\cite{Ozulker:2025ehg}, full CPL has the lowest sampled $\chi^2$ in all six data combinations; in two of them the improvement over CPL$_{<a_c}$ is $\Delta\chi^2=1.4$.

\subsection{The fold and the adiabatic sound speed}
\label{sec:fold}
The double valuedness is a property of $\rde$ as independent variable. In terms of $\nu\equiv(a_I/a)^3$, the ratio of a conserved number density to its value at the fold,
\begin{equation}
 \rde=\rho_I\,\nu^{s}\exp\bigl[3s\,(\nu^{-1/3}-1)\bigr]
 \label{eq:rhon}
\end{equation}
is single valued and analytic for all $\nu>0$, with $\pde=\nu\,\dd\rde/\dd\nu-\rde$. This homogeneous perfect-fluid construction passes through $w=-1$, where its enthalpy per particle, proportional to $\dd\rde/\dd\nu$, changes sign. It is not a globally single-valued closure $p=p(\rho)$. Nor does its regularity in $\nu$ guarantee regular perturbations: the derivative along the background diverges at the fold,
\begin{equation}
 \frac{\dd\pde}{\dd\rde}=w-\frac{W_k}{3(1+W_k)}=\frac{3w^2+2w+w_i}{3(1+w)},
 \label{eq:slope}
\end{equation}
with $W_k=W_k(Z)$. The first form gives the slope as a function of density; the numerator of the second is the polynomial used in Ref.~\cite{Perkovic:2020mph} to determine where it is nonnegative. Near any crossing fold, $\dd\pde/\dd\rde\simeq1/(3\Uc_k)$: it tends to $+\infty$ on $k=0$ and $-\infty$ on $k=-1$. Away from the fold it vanishes at physically attained roots $w=[-1\pm\sqrt{1-3w_i}\,]/3$, where the pressure is extremal and the history meets $w'=3w(1+w)$ \cite{dePutter:2007ny}.

For a literal adiabatic barotropic perfect fluid, Eq.~\eqref{eq:slope} is the physical sound speed squared, $c_s^2=\dd p/\dd\rho$. Its positive divergence and its negative values on the outgoing side obstruct a regular, stable, causal fluid realization near the fold \cite{Kunz:2006wc,Perkovic:2020mph}. For an effective CPL background with independent entropy perturbations or additional fields, Eq.~\eqref{eq:slope} specifies only the adiabatic derivative $c_a^2=p'/\rho'$; the physical propagation speed requires a separate perturbative closure. A regular single-valued barotrope cannot execute this crossing, and a canonical minimally coupled scalar cannot enter the phantom regime, but the perfect-fluid form of the homogeneous stress tensor alone imposes neither restriction \cite{Vikman:2004dc,Hu:2004kh,Kunz:2006wc}. The branch constructions above should therefore be read as background solutions, without assuming a stable microscopic completion.

\section{A log-concavity theorem for cosmic acceleration}
\label{sec:theorem}
\subsection{Source ratios}
We now turn from the fluid to the expansion; nothing in this section uses the closed form of Sec.~\ref{sec:eos}. We assume general relativity on an FLRW background of arbitrary spatial curvature, filled with separately conserved components labeled $j$. Then
\begin{equation}
 \frac{\ddot a}{a}=-\frac{4\pi G}{3}\,\Mtot,\qquad \Mtot\equiv\sum_j(\rho_j+3p_j),
 \label{eq:raych}
\end{equation}
in which curvature does not appear. The Friedmann constraint nevertheless determines which scale factors an expanding solution can reach. We distinguish the source functions, which may be defined on a larger interval of $a$, from a connected expanding branch with $H^2>0$. We single out one component, labeled ``de,'' with $\rde>0$ and equation of state $w(N)$, and write $\Mbg\equiv\Mtot-\Mde$ and $\Ibg\equiv\Itot-\Ide$ for the active and inertial densities of all the others, both assumed positive. Define
\begin{equation}
 K\equiv-\frac{\Mde}{\Mbg},\qquad J\equiv-\frac{\Ide}{\Ibg}.
 \label{eq:ratios}
\end{equation}
Then $\Mtot<0$ if and only if $K>1$, and $\Itot<0$ if and only if $J>1$. On a physical FLRW solution these are respectively the conditions for acceleration and total NEC violation. Both statements are about superlevel sets of a single function, and $K$ ($J$) is positive only where the component is repulsive (NEC violating).

\begin{lemma}
\label{lem:identity}
Where $w<-\tfrac13$, with $\epsilon\equiv-(1+3w)>0$,
\begin{align}
 (\ln K)'&=\epsilon-2-\frac{3w'}{\epsilon}-(\ln\Mbg)',\label{eq:lnK1}\\
 (\ln K)''&=-3w'-\frac{3w''}{\epsilon}-\frac{9w'^2}{\epsilon^2}-(\ln\Mbg)''.\label{eq:lnK2}
\end{align}
Where $w<-1$, with $\eta\equiv-(1+w)>0$,
\begin{align}
 (\ln J)'&=3\eta-\frac{w'}{\eta}-(\ln\Ibg)',\label{eq:lnJ1}\\
 (\ln J)''&=-3w'-\frac{w''}{\eta}-\frac{w'^2}{\eta^2}-(\ln\Ibg)''.\label{eq:lnJ2}
\end{align}
\end{lemma}
\begin{proof}
$\ln K=\ln\rde+\ln\epsilon-\ln\Mbg$ and $\ln J=\ln\rde+\ln\eta-\ln\Ibg$, with $(\ln\rde)'=-3(1+w)$.
\end{proof}

\begin{lemma}
\label{lem:background}
Let the remaining content consist of (a)~positive-density fluids of constant $w_j>-\tfrac13$, (b)~collisionless massive species with nonnegative distribution functions, and optionally (c)~a negative cosmological constant, which contributes a positive active density. Assume $\Mbg>0$ and $\Ibg>0$. Then $\ln\Mbg$ is a convex function of $N$, and without (c) also $(\ln\Mbg)'\le-2$. For $\Ibg$ the same holds with $w_j>-1$ in (a) and $(\ln\Ibg)'<0$; a cosmological constant of either sign does not contribute to $\Ibg$.
\end{lemma}
\begin{proof}
A sum of log-convex functions is log-convex \cite{Boyd:2004fnq}. The terms of type (a) are $\propto e^{-3(1+w_j)N}$ and log-linear; that of type (c) is the constant $-2\rho_\Lambda>0$; those of type (b) are log-convex with logarithmic slope in $[-4,-3]$, including the dust limit (Appendix~\ref{app:nu}). The logarithmic slope of a sum is a weighted mean of the slopes of its terms.
\end{proof}

\subsection{The theorem}
\begin{theorem}
\label{thm:single}
Let a component with $\rde>0$ be separately conserved with $w$ twice differentiable on an interval $S$ of $N=\ln a$, and let the remaining content be as in Lemma~\ref{lem:background}.

(a) If $w'\ge0$ and $w''\ge0$ on $S$, the source-sign set $\{N\in S:\Mtot<0\}$ is an interval (possibly empty). So is $\{N\in S:\Itot<0\}$.

(b) If $w'\le0$ on $S$ and there is no negative cosmological constant, each of these sets is empty or a final segment of $S$.

On a connected expanding branch, restriction to its physical domain within $S$ gives the accelerating and total-NEC-violating epochs; their connectedness is preserved, and in case (b) neither ends once begun within that domain. In both cases $\Itot<0$ implies $\Mtot<0$ provided $\rho_{\rm tot}>0$.
\end{theorem}
\begin{proof}
(a) Since $w$ is nondecreasing, the part $S_1$ of $S$ on which $w<-\tfrac13$ is an initial segment. On it every term on the right of Eq.~\eqref{eq:lnK2} is nonpositive, the last by Lemma~\ref{lem:background}, so $\ln K$ is concave and $\{K>1\}=\{\ln K>0\}$ is an interval. Outside $S_1$, $K\le0$. The argument for $J$ on $\{w<-1\}$ is identical, with Eq.~\eqref{eq:lnJ2}. (b) Now $\{w<-\tfrac13\}$ is a final segment of $S$, on which Eq.~\eqref{eq:lnK1} and Lemma~\ref{lem:background} give $(\ln K)'\ge\epsilon>0$. Hence $K$ increases and $\{K>1\}$ is a final segment; likewise $(\ln J)'\ge3\eta>0$. The last statement follows from $\Mtot=3\Itot-2\rho_{\rm tot}$.
\end{proof}

In case (a), $K$ is log-concave on its positive domain and cannot exceed unity in separated intervals. In case (b), it increases there. These are distinct mechanisms for the same connectedness restriction.

\subsection{Scope and sharpness}
\emph{Curvature and matter content.} Curvature is absent from Eq.~\eqref{eq:raych}, so the theorem applies to the source-sign sets whatever the sign of the curvature. Their intersection with the connected domain of an expanding solution gives at most one interval of time. Curvature can, however, prevent that solution from reaching a formal onset or exit.\footnote{For example, dust plus a positive cosmological constant with $(\Omega_{m0},\Omega_{\Lambda0},\Omega_{k0})=(2.06,0.01,-1.07)$ has $a^3H^2/H_0^2=0.01(a-2)(a^2+2a-103)$. Its expanding Big-Bang branch turns at $a=2$, before the formal acceleration onset $a=103^{1/3}$.} The admissible content is broader than dust and radiation, and includes massive neutrinos \cite{Lesgourgues:2006nd}. Case (b) extends to several components, each with a nonincreasing equation of state, a positive cosmological constant among them: with $F\equiv-\sum(\rho+3p)/\Mbg$, the sum running over these components, one has $F'\ge\sum[(1+3w)^2-3w']\rho/\Mbg\ge0$ wherever $F>0$, so $\{F>1\}$ is again a final segment.

\emph{Locality.} $S$ need not be the whole history. Where a parametrization is trusted only over a range of redshift, the theorem applies on that range.

\emph{Which equations of state.} CPL has $w'=w-w_i=-w_aa$ and $w''=w'$: case (a) for $w_a\le0$ and case (b) for $w_a\ge0$. The same alternatives hold for $w=w_i-w_na^n$ with $n>0$, to which the closed form of Sec.~\ref{sec:explicit} extends, and for any $w$ linear in $\ln a$. Thawing histories, $w'>0$ \cite{Caldwell:2005tm}, satisfy (a) wherever they are convex in $\ln a$. By Eq.~\eqref{eq:lnK2}, the acceleration conclusion of (a) also follows if $\{w<-\tfrac13\}$ is an interval on which $w''+\epsilon w'+3w'^2/\epsilon\ge0$. For total NEC violation, Eq.~\eqref{eq:lnJ2} gives the analogous sufficient condition: $\{w<-1\}$ is an interval on which $w''+3\eta w'+w'^2/\eta\ge0$.

\begin{figure}[t]
 \centering
 \includegraphics[width=\columnwidth]{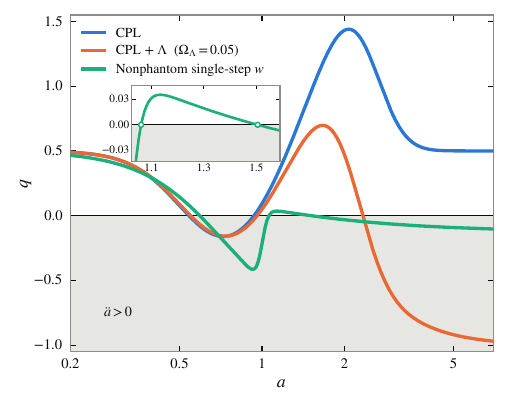}
 \caption{Deceleration parameter for three histories; shading under the axis marks $\ddot a>0$. The reference history (Table~\ref{tab:desi}, BAO+CMB) accelerates once. Replacing part of its CPL density by a cosmological constant with $\Omega_\Lambda=0.05$ produces two accelerating epochs, as does dust plus the monotonic, nonphantom equation of state~\eqref{eq:single-step}. The inset resolves the latter history's shallow decelerating interval. The two counterexamples each violate one hypothesis of Theorem~\ref{thm:single}.}
 \label{fig:sharp}
\end{figure}

\emph{Sharpness.} Three examples, each violating one hypothesis, have two accelerating epochs; the first two are shown in Fig.~\ref{fig:sharp}. (1)~\emph{A second repulsive component.} Replacing part of the CPL density of the reference history by a cosmological constant with $\Omega_\Lambda=0.05$ gives $\ddot a>0$ for $0.548<a<0.960$ and again for $a>2.33$. (2)~\emph{Monotonic but nonconvex $w$.} For flat dust plus dark energy with $\Om=0.3$, take
\begin{equation}
 w(N)=-0.95+\tfrac{8}{15}\,\sigma(40N),
 \label{eq:single-step}
\end{equation}
where $\sigma(v)\equiv(1+e^{-v})^{-1}$ is the logistic function. Here $w$ increases from $-19/20$ to $-5/12$, so $-1<w<-\tfrac13$ throughout; $w'>0$, while $w''<0$ for $N>0$. Conservation gives
\begin{equation}
 K(N)=\frac73 e^{2.85N}
 \left(\frac{1+e^{40N}}2\right)^{-1/25}(1.85-1.6\sigma),
 \label{eq:single-step-K}
\end{equation}
where $\sigma$ stands for $\sigma(40N)$. Writing $r=\rde/\rho_m$, one finds $K'/r=66.56\sigma^2-71.52\sigma+5.2725$, with exactly two roots in $(0,1)$. Together with $K(-\infty)=0$, $K(\infty)=\infty$ and $K(-0.15)\simeq2.888>1>K(0.15)\simeq0.735$, this establishes exactly three crossings of unity. Acceleration occurs for $0.593\lesssim a\lesssim1.061$ and again for $a\gtrsim1.505$. Monotonicity alone therefore does not suffice, even for a component that remains nonphantom and repulsive.

(3)~\emph{Case (b) with a negative cosmological constant.} For a flat universe with $\Om=0.3$, $\Omega_\Lambda=-0.05$, and $w$ falling from $-\tfrac12$ to $-2$ in a smooth step of width $0.05$ in $\ln a$ at $a=4$, the expansion accelerates for $1.10<a<1.89$ and again from $a=3.58$, with $\rho_{\rm tot}>0$ throughout.

\emph{Relation to sign-switching dark energy.} Two separated accelerating epochs occur in the sign-switching backgrounds of Ref.~\cite{Akarsu:2026pia}, which are of the kind studied in Refs.~\cite{Akarsu:2019hmw,Akarsu:2021fol,Akarsu:2022typ}; Sec.~IV\,B of Ref.~\cite{Akarsu:2026pia} bounds the number of transitions under assumptions on the stationary points of the total source in flat dust backgrounds. Examples (1) and (2) show that a change of sign of the density is not needed: an additional repulsive component, or a loss of convexity of $w(\ln a)$, suffices. Theorem~\ref{thm:single} instead controls the number of epochs through conditions on the equation of state, and holds with radiation, collisionless massive species and curvature.

\section{Acceleration in CPL: thresholds and windows}
\label{sec:cpl}
\subsection{One epoch, three cases}
\begin{corollary}
\label{cor:cpl}
For positive-density CPL and a nonzero background consisting of dust, radiation and/or collisionless massive species, define the source-sign sets $\mathcal A\equiv\{a>0:\Mtot(a)<0\}$ and $\mathcal P\equiv\{a>0:\Itot(a)<0\}$. For any spatial curvature:

(i) If $w_a>0$, $\mathcal A=(a_{\rm acc},\infty)$ and $\mathcal P=(a_{\rm NEC},\infty)$, with $a_{\rm acc}<a_{\rm NEC}$.

(ii) If $w_a=0$, $\mathcal A$ is a final interval with a finite onset if $w_0<-\tfrac13$, and $\mathcal P$ is such an interval if $w_0<-1$; otherwise the corresponding set is empty.

(iii) If $w_a<0$, $\mathcal A$ is empty or a bounded interval $(a_-,a_+)$ with $a_+<a_M$; it is empty when $w_i\ge-\tfrac13$. A nonempty $\mathcal P$ requires $w_i<-1$ and is a subinterval of $(a_-,a_+)$ that ends before $a_I$.

An expanding solution realizes the part of each set that lies in its domain of scale factors; flat and open models realize the full classification.
\end{corollary}
\begin{proof}
The source functions are defined for all $a>0$, so Theorem~\ref{thm:single} applies on $S=\mathbb R$. The ordinary background dilutes at least as fast as dust. Thus $K\to0$ as $a\to0$ if the component is initially repulsive, and similarly $J\to0$ if it is initially NEC violating; otherwise the respective positive domain begins at $K=0$ or $J=0$. For $w_a>0$, the factor $e^{3w_aa}$ in $\rde$ gives $K,J\to\infty$ at large $a$. The constant-$w$ limits follow from the same background dilution bound. For $w_a<0$, $K>0$ only for $a<a_M$, and $J>0$ only for $a<a_I$, giving bounded positive domains. Since all component densities are positive, total NEC violation lies strictly inside the accelerating source-sign interval whenever its boundary is finite.
\end{proof}

Under the assumptions of Corollary~\ref{cor:cpl}, CPL permits at most one accelerating epoch on a connected expanding branch. For $w_a<0$, continuation of the ansatz gives an empty or bounded accelerating window in scale factor; observations constrain the model parameters and hence the window's existence and location.

\subsection{Dust limit}
Neglecting radiation, whose effect is quantified below, $\Mbg=\Ibg=\rho_m\propto a^{-3}$. Let
\begin{equation}
 \begin{gathered}
 \alpha\equiv-3w_i,\qquad\beta\equiv3w_a,\qquad y\equiv-3w=\alpha+\beta a,\\
 r\equiv\frac{\rde}{\rho_m}=r_0\,a^{\alpha}e^{\beta(a-1)},\qquad r_0=\frac{\Ode}{\Om},
 \end{gathered}
 \label{eq:ratio}
\end{equation}
with no assumption on curvature. Then
\begin{equation}
 K=r\,(y-1),\qquad J=\tfrac13\,r\,(y-3),
 \label{eq:KJ}
\end{equation}
and, because $y'=y-\alpha$ on the CPL line,
\begin{equation}
 K'=r\,(y^2-\alpha),\qquad J'=\tfrac13\,r\,(y^2-2y-\alpha).
 \label{eq:Kprime}
\end{equation}
For $w_a<0$ and $\alpha>1$ the ratio $K$ therefore peaks at $y=\sqrt\alpha$, that is, when the equation of state reaches
\begin{equation}
 \begin{gathered}
 w_K=-\sqrt{-w_i/3},\\
 a_K=\frac{\alpha-\sqrt\alpha}{|\beta|}=\frac{\sqrt\alpha}{\sqrt\alpha+1}\,a_M ,
 \end{gathered}
 \label{eq:peak}
\end{equation}
an epoch that does not depend on the densities. It is the peak of $K$, not the minimum of $q$.

Equations~\eqref{eq:Kprime} are instances of two identities that hold for any separately conserved component, $[a^3(\rho+3p)]'=3a\,(a^2p)'$ and $[a^3(\rho+p)]'=a^3p'$. With dust they give
\begin{equation}
 K'=-\frac{3a}{\rho_{m0}}\,(a^2\pde)',\qquad J'=-\frac{a^3}{\rho_{m0}}\,\pde' .
 \label{eq:pressureid}
\end{equation}
The source ratio $K$ grows for as long as $a^2\pde$ decreases, and $J$ for as long as the pressure does. In the pressure--density plane of Fig.~\ref{fig:eos}(a), $J$ therefore peaks at the lowest point of the trajectory, where the slope~\eqref{eq:slope} vanishes. The maximum of $K$ instead satisfies $(a^2\pde)'=0$, or $2\pde+\pde'=0$. Where $\rde'\ne0$, this is equivalently $\dd\pde/\dd\rde=2w/[3(1+w)]$.

The whole profile has a scaling form. In terms of $\xi\equiv|\beta|a$,
\begin{equation}
 \begin{gathered}
 K=\frac{\varphi_\alpha(\xi)}{\kappa},\qquad\varphi_\alpha(\xi)\equiv\xi^{\alpha}e^{-\xi}(\alpha-1-\xi),\\
 \kappa\equiv|\beta|^{\alpha}e^{-|\beta|}\,\frac{\Om}{\Ode}.
 \end{gathered}
 \label{eq:profile}
\end{equation}
The shape of $K$ as a function of $\ln a$ depends on $\alpha$ alone; $w_a$ and the densities enter only through the position $\xi=|\beta|a$ and the single number $\kappa$. With
\begin{equation}
 \phihat(\alpha)\equiv\max_\xi\varphi_\alpha=(\sqrt\alpha-1)(\alpha-\sqrt\alpha)^{\alpha}e^{\sqrt\alpha-\alpha},
 \label{eq:phihat}
\end{equation}
the peak value is $\Kmax=\phihat(\alpha)/\kappa$, and the accelerating window exists if and only if $\Kmax>1$:
\begin{equation}
 \boxed{\frac{\Ode}{\Om}>\frac{|\beta|^{\alpha}e^{-|\beta|}}{\phihat(\alpha)}\,.}
 \label{eq:threshold}
\end{equation}
For a flat universe this reads $\Om<\Omega_{m0,\rm crit}\equiv[1+|\beta|^{\alpha}e^{-|\beta|}/\phihat(\alpha)]^{-1}$. At equality $\Mtot$ touches zero at $a_K$ without becoming negative; for $\Kmax>1$ its two zeros delimit the window. In a closed model the expanding solution must also reach it.

In class~IV the threshold can be written in the variables of the fluid. There $\xi=3|s|\,x$ and $\alpha=3(1+|s|)$, so that Eq.~\eqref{eq:profile} becomes $K=r_I\,x^{\alpha}e^{3|s|(1-x)}\,[\,2+3|s|(1-x)]$, where $r_I\equiv\rho_I/\rho_m(a_I)$ is the ratio of the dark-energy to the matter density at the fold, and
\begin{equation}
 \Kmax=\frac{r_I}{r_{\rm crit}(s)},\qquad r_{\rm crit}(s)\equiv\frac{(3|s|)^{\alpha}e^{-3|s|}}{\phihat(\alpha)} .
 \label{eq:foldthreshold}
\end{equation}
Equivalently, the window exists if and only if the dark-energy density at its maximum exceeds $r_{\rm crit}(s)$ times the matter density at that epoch. The critical ratio depends on the equation of state through $s$ alone, and the profile of $K$ in $a/a_I$ through $s$ and $r_I$; the third parameter decides only where the present falls. At the fold itself $K=2r_I$. For the flat histories of Table~\ref{tab:desi}, $r_{\rm crit}=0.25$--$0.46$ and $r_I=0.76$--$0.96$: these histories accelerate at the fold although matter still dominates there.

\begin{table*}[t]
\caption{Flat dust-plus-CPL histories at the DESI DR2 mean values \cite{DESI:2025zgx} (Table~V there), each with its own jointly inferred $\Om$. The fold occurs at $a_I$ with density $\rho_I$; $\Kmax>1$ gives an accelerating window $(a_-,a_+)$, and $\Omega_{m0,\rm crit}$ is the threshold~\eqref{eq:threshold}. The present status follows from $q_0$, while $\Jmax<1$ excludes total NEC violation. All four histories belong to class~IV. The derived quantities are evaluated at these means; they are not posterior means or credible intervals.}
\label{tab:desi}
\begin{ruledtabular}
\begin{tabular}{lccccccccccc}
Data & $w_0$ & $w_a$ & $\Om$ & $a_I$ & $\rho_I/\rho_{{\rm de}0}$ & $\Kmax$ & $\Omega_{m0,\rm crit}$ & $a_-$ & $a_+$ & $q_0$ & $\Jmax$\\
\hline
BAO+CMB & $-0.42$ & $-1.75$ & $0.353$ & $0.669$ & $1.387$ & $1.65$ & $0.473$ & $0.541$ & $0.935$ & $+0.092$ & $0.084$\\
\quad +PantheonPlus & $-0.838$ & $-0.62$ & $0.3114$ & $0.739$ & $1.072$ & $3.87$ & $0.637$ & $0.575$ & $1.728$ & $-0.366$ & $0.044$\\
\quad +Union3 & $-0.667$ & $-1.09$ & $0.3275$ & $0.694$ & $1.187$ & $2.17$ & $0.514$ & $0.561$ & $1.196$ & $-0.173$ & $0.060$\\
\quad +DESY5 & $-0.752$ & $-0.86$ & $0.3191$ & $0.712$ & $1.127$ & $2.69$ & $0.557$ & $0.568$ & $1.385$ & $-0.268$ & $0.053$\\
\end{tabular}
\end{ruledtabular}
\end{table*}

\subsection{The accelerating window}
Because Eq.~\eqref{eq:profile} involves $a$ only through $\xi$, the edges of the window satisfy $\varphi_\alpha(\xi_\pm)=\phihat(\alpha)/\Kmax$. Its e-fold length,
\begin{equation}
 \Delta N_{\rm acc}\equiv\ln\frac{a_+}{a_-}=\ln\frac{\xi_+}{\xi_-},
 \label{eq:duration}
\end{equation}
is thus an exact function of two numbers only, $\alpha=-3(w_0+w_a)$ and $\Kmax$. At fixed $\alpha>1$, expanding $\ln K$ about its maximum (Appendix~\ref{app:peak}), with
\begin{equation}
 \lambda\equiv\ln\Kmax,\qquad t\equiv\sqrt{\lambda/\alpha},
\end{equation}
gives the cubic near-threshold approximation
\begin{equation}
 \ln\frac{a_\pm}{a_K}=\pm t-\frac{\sqrt\alpha+2}{6}\,t^2\pm\frac{4+\sqrt\alpha-2\alpha}{36}\,t^3+O(t^4),
 \label{eq:window}
\end{equation}
\begin{equation}
 \Delta N_{\rm acc}=2t\Bigl[1-\frac{2\alpha-\sqrt\alpha-4}{36}\,t^2\Bigr]+O(t^5).
 \label{eq:width}
\end{equation}
The remainders are asymptotic as $t\to0$ at fixed $\alpha$. The window opens as a square root, as at any nondegenerate maximum, with a coefficient fixed by $(\ln K)''=-2\alpha$ at the peak; it is skewed toward early times by the $t^2$ term. The approximation also performs well at the four DESI DR2 mean values of Table~\ref{tab:desi}, where $\Kmax$ is between $1.6$ and $3.9$: Eq.~\eqref{eq:window} reproduces both edges to better than $1\%$, and to $5\times10^{-4}$ for BAO+CMB. For BAO+CMB, the leading width term $2t$ alone overestimates $\Delta N_{\rm acc}$ by $1.2\%$.

\emph{Radiation.} At fixed present matter and dark-energy densities, adding radiation gives $\Mbg=\rho_m(1+c_r/a)$ with $c_r\equiv2\rho_{r0}/\rho_{m0}=2\Omega_{r0}/\Om$, so $K\to Ka/(a+c_r)$ and, to first order in $c_r$,
\begin{equation}
 \ln\Kmax\;\to\;\ln\Kmax-\frac{c_r}{a_K}.
 \label{eq:radiation}
\end{equation}
For the reference history and $\Omega_{r0}=9.0\times10^{-5}$, $c_r\simeq5.1\times10^{-4}$: $\Kmax$ decreases by $0.068\%$, while the onset and exit shift by approximately $+2.0\times10^{-4}$ and $-0.8\times10^{-4}$ in $a$, respectively.

\emph{Total NEC.} For $w_a<0$ and $\alpha>3$, $J$ peaks at $y=1+\sqrt{\alpha+1}$, i.e., at
\begin{equation}
 \begin{gathered}
 w_J=-\tfrac13\bigl(1+\sqrt{1-3w_i}\bigr),\\
 a_J=\frac{\alpha-1-\sqrt{\alpha+1}}{|\beta|},
 \end{gathered}
\end{equation}
with $\Jmax=\psihat(\alpha)/\kappa$ and
\begin{equation}
 \psihat(\alpha)\equiv\tfrac13(\sqrt{\alpha+1}-2)\,(\alpha-1-\sqrt{\alpha+1})^{\alpha}\,e^{1+\sqrt{\alpha+1}-\alpha}.
 \label{eq:psihat}
\end{equation}
The total-NEC-violating set is nonempty if and only if $\Jmax>1$, a stronger requirement than Eq.~\eqref{eq:threshold} since $\psihat<\phihat$.

\begin{figure*}[t]
 \centering
 \includegraphics[width=\textwidth]{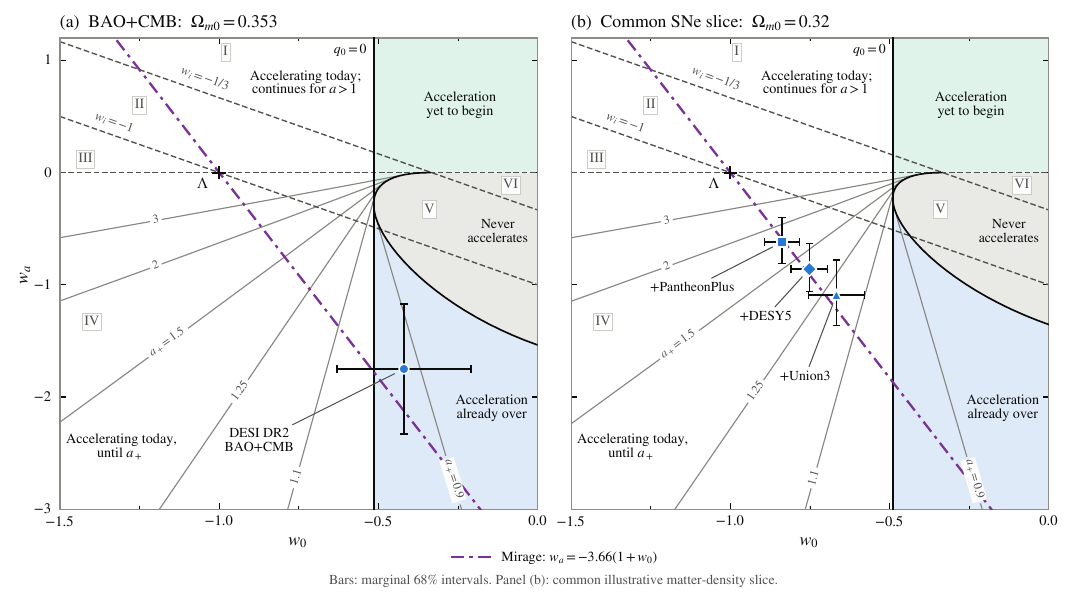}
 \caption{The $(w_0,w_a)$ plane for spatially flat dust plus CPL. Dashed lines separate the six classes of Table~\ref{tab:sectors} (roman numerals). The fills give the present status of cosmic acceleration; their boundaries are the line $q_0=0$, $w_0=-1/(3\Ode)$, and the two branches of the threshold curve $\Kmax=1$, Eq.~\eqref{eq:threshold}, which meet it at the point~\eqref{eq:triple}. Thin contours give the exit epoch $a_+$ of the accelerating window where it exists. Panel (a) uses $\Om=0.353$ for BAO+CMB; panel (b) is a common $\Om=0.32$ slice for the combinations with the PantheonPlus, Union3 and DESY5 supernova samples. Symbols show the DESI DR2 mean values \cite{DESI:2025zgx}; horizontal and vertical bars are marginal $68\%$ intervals (the joint posteriors, elongated along the anticorrelation direction, are not drawn). Table~\ref{tab:desi} evaluates each combination at its own jointly inferred matter density. The cross marks a cosmological constant. The purple dash--dotted line is the conventional mirage calibration $w_a=-3.66(1+w_0)$ of Sec.~\ref{sec:plane}.}
 \label{fig:map}
\end{figure*}

\begin{figure*}[t]
 \centering
 \includegraphics[width=\textwidth]{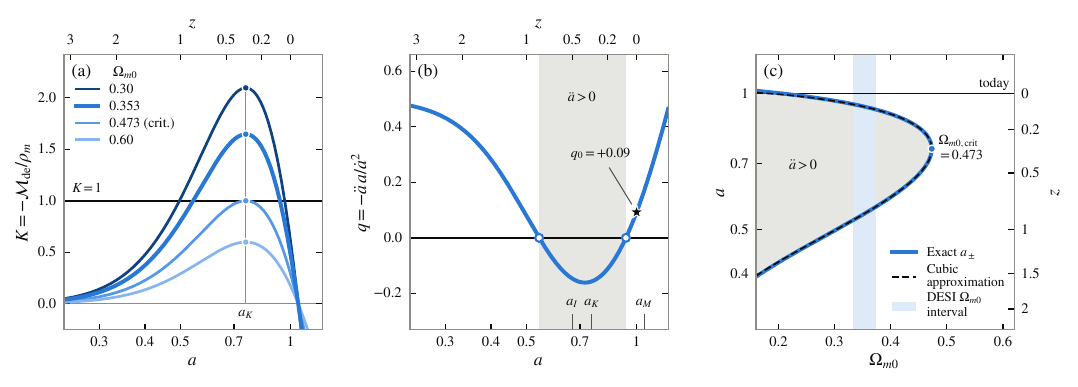}
 \caption{The accelerating window of the reference history $(w_0,w_a)=(-0.42,-1.75)$, in the dust limit. (a)~The source ratio $K$ for four values of $\Om$; acceleration requires $K>1$, and the peak epoch $a_K$ is the same for all. (b)~Deceleration parameter for $\Om=0.353$; the band is the window~\eqref{eq:accwindow}, the star is $a=1$, and the ticks mark the NEC crossing $a_I$, the peak $a_K$, and the SEC crossing $a_M$. (c)~The window as a function of $\Om$: it closes at $\Omega_{m0,\rm crit}=0.473$ and, when it exists, ends before today for $\Om>0.206$. Dashed lines are the cubic endpoint approximation~\eqref{eq:window}; the vertical band shows the marginal DESI BAO+CMB interval $\Om=0.353\pm0.021$ at fixed $(w_0,w_a)$.}
 \label{fig:acceleration}
\end{figure*}

\section{\texorpdfstring{The $(w_0,w_a)$ plane and the DESI quadrant}{The (w0, wa) plane and the DESI quadrant}}
\label{sec:plane}
\subsection{Present status of acceleration}

For a flat universe with negligible radiation, where $a=1$ lies relative to the window follows from $K(1)=r_0(-1-3w_0)$ and from the position of the peak. The expansion accelerates today if and only if
\begin{equation}
 w_0<w_\times\equiv-\frac{1}{3\Ode},
 \label{eq:q0line}
\end{equation}
since $q_0=\tfrac12(1+3w_0\Ode)$. For $w_a<0$ and $w_i\ge-\tfrac13$ acceleration never occurs. For $w_0>w_\times$, $w_a<0$ and $w_i<-\tfrac13$, there are three possibilities: acceleration is over ($a_+<1$) if $\Kmax>1$ and $w_0>w_K$; it has yet to begin ($a_->1$) if $\Kmax>1$ and $w_0<w_K$; and it never occurs if $\Kmax\le1$. For $w_0>w_\times$ and $w_a>0$ it has yet to begin. The three boundaries meet at
\begin{equation}
 (w_0,w_a)=\bigl(w_\times,\,-3w_\times^2-w_\times\bigr),
 \label{eq:triple}
\end{equation}
where $a_K=1$ and $\Kmax=1$. Figure~\ref{fig:map} combines this map with the six classes and contours of $a_+$.

\emph{Mirage dark energy.} The calibrated relation $w_a=-\mu(1+w_0)$, conventionally $\mu=3.66$, selects a one-parameter CPL family motivated by approximate matching of the $\Lambda$CDM distance to CMB last scattering \cite{Linder:2007ka,DESI:2024kob}. For fixed $\mu>1$ and $w_0\ne-1$, one has $s=-(\mu-1)(1+w_0)$ and $a_I=1-\mu^{-1}$, giving $a_I\simeq0.727$ ($z_I\simeq0.376$) for this calibration. The $w_0>-1$ ray lies in class~IV, ${\rm P}\to{\rm R}\to{\rm A}$, with a density maximum; any acceleration is a bounded window (Corollary~\ref{cor:cpl}). The opposite ray has a density minimum: class~II for $-1-2/[3(\mu-1)]\le w_0<-1$, and class~I below this range. The point $(-1,0)$ is the cosmological-constant limit. The same calibration fixes today's normalized position, $u_0=-1/(\mu-1)$ and $L_0=1/(\mu-1)+\ln(1-1/\mu)$, on $W_{-1}$; it does not fix the acceleration endpoints. Figure~\ref{fig:map} shows the conventional fixed-slope line in both panels. Its distance-matching interpretation is approximate, with a weak background dependence of $\mu$ \cite{DESI:2024kob}.

\subsection{The DESI DR2 mean values}
\label{sec:desi}

Table~\ref{tab:desi} lists BAO+CMB and each of the three SN combinations presented in Ref.~\cite{DESI:2025zgx}. It shows where the inferred parameters place the present within the history that CPL allows. All four mean values lie in class~IV, with a density maximum at $a_I=0.67$--$0.74$ and $\rho_I/\rho_{{\rm de}0}=1.07$--$1.39$.

For the reference history the accelerating window is
\begin{equation}
 0.541<a<0.935,\qquad 0.069<z<0.847,
 \label{eq:accwindow}
\end{equation}
so today lies beyond it. The three SN-inclusive histories are still accelerating today. All four histories cross the dark-energy NEC boundary, while $\Jmax<0.09$ ensures that the total fluid satisfies the NEC throughout. Caldwell and Linder illustrate this distinction for representative histories \cite{Caldwell:2025inn}; Xu et al.\ map the total-fluid NEC region alongside reconstructions of $q(z)$ \cite{Xu:2026sbw}. These point evaluations do not give the probability of either present status: the BAO+CMB posterior of $q_0$ straddles zero \cite{Colgain:2025nzf,Wang:2025owe,Xu:2026sbw,Ferri:2026sat}. Corollary~\ref{cor:cpl} supplies the common structural restriction; the data locate the present within it.

Figure~\ref{fig:acceleration} follows the reference history in detail. The $K$ peak at $a_K=0.754$ occurs after the minimum of $q$ at $a=0.722$. Acceleration ends at $a_+=0.935$, while the component remains repulsive until $a_M=1.050$; the ordering $a_+<a_M$ follows generally from Corollary~\ref{cor:cpl}. In Fig.~\ref{fig:eos}(a), the accelerating arc enters on $W_0$ at $\rde=0.930\rho_I$, traverses the fold, and exits on $W_{-1}$ at $0.801\rho_I$. Thus this history accelerates only within $20\%$ of its maximal dark-energy density. The SN-inclusive arcs also span the fold, but exit at $0.51$--$0.67\rho_I$.

At fixed $(w_0,w_a)=(-0.42,-1.75)$, the window exists and has ended today for $0.20635<\Om<0.47316$ [Fig.~\ref{fig:acceleration}(c)]. Taking $\Om=0.30$ instead gives $0.038<z<1.013$, reproducing the example in Ref.~\cite{Akarsu:2026pia}. The source geometry fixes the allowed sequence; the relative matter abundance fixes whether and where acceleration occurs.

The degree of extrapolation also matters. For BAO+CMB, the NEC crossing ($z_I=0.50$) and $K$ peak ($z=0.33$) lie in the BAO redshift range, whereas the exit at $z=0.07$ lies below its lowest bin, $0.1<z<0.4$ \cite{DESI:2025zgx}. Reference~\cite{Ferri:2026sat} links the nonaccelerating present to the absence of BAO data below $z_{\rm eff}\simeq0.3$. The SN-inclusive exits, $a_+=1.2$--$1.7$, follow from extrapolating CPL. For example, the Barboza--Alcaniz form $w(a)=w_0+w_a(1-a)/[a^2+(1-a)^2]$ \cite{Barboza:2008rh} matches CPL's present value and first derivative, but tends to $w_0$ as $a\to\infty$. For $w_0<-\tfrac13$ it gives persistent late-time acceleration in flat dust plus positive-density dark energy, although it need not fit the data equally well over the observed range. At the reference values its first accelerating window, $0.538<a<0.943$, nearly coincides with the CPL window~\eqref{eq:accwindow}, but acceleration formally resumes at $a\simeq15.4$ and, since $w$ decreases for $a>1+1/\sqrt2$, never ends by case~(b) of Theorem~\ref{thm:single}. The single bounded window is thus imposed by the CPL form, not by today's value and slope of $w$. Theorem~\ref{thm:single} may instead be restricted to the range over which a parametrization is trusted.

\section{Conclusions}
We asked what the CPL parametrization imposes independently of the data. The first answer concerns its equation of state. The pressure--density relation has the closed form~\eqref{eq:eos}, which resolves its real Lambert branches and their physical domains, and its positive-density histories with $w_a\ne0$ fall into six classes. Every history that crosses $w=-1$ reduces, in normalized variables, to the parameter-free curve~\eqref{eq:universal}, folded at the density extremum: the equation of state has two parameters, while an integration constant locates the present. This global organization extends the earlier implicit and graphical descriptions.

The fold separates a prescribed background from an autonomous barotropic closure. Each branch, taken as a fluid in its own right, ends on or starts from a constant-density plateau; for a specified nonconstant incoming history, uninterrupted CPL is the unique crossing continuation with differentiable $w(a)$. The divergent adiabatic derivative at the fold remains an obstruction to a regular barotropic realization, and a viable perturbative completion is additional input.

The second answer concerns the expansion. On a connected expanding branch, Theorem~\ref{thm:single} allows at most one accelerating epoch, and at most one epoch of total NEC violation, for a separately conserved, positive-density component whose equation of state is nondecreasing and convex in $\ln a$, or nonincreasing, accompanied by dust, radiation and collisionless massive species, with any spatial curvature. Counterexamples show that a second repulsive component, or a monotonic but nonconvex $w$, already permits two separated accelerating epochs.

For CPL with $w_a<0$, the case relevant to the DESI mean values, any acceleration is therefore confined to a single bounded window. In the dust limit the window exists if and only if the threshold~\eqref{eq:threshold} is met, and Eq.~\eqref{eq:window} approximates its edges. In class~IV the threshold links the two parts of this work: the window exists if and only if the dark-energy-to-matter density ratio at the fold exceeds a critical value fixed by $s=1+w_0+w_a$ alone, Eq.~\eqref{eq:foldthreshold}. The four DESI DR2 mean values all lie in class~IV and involve no total NEC violation; they place the present beyond the window at the BAO+CMB mean, where it closes at $z\simeq0.07$, and inside it for the three combinations that include supernovae. This confinement assumes the stated sources and continued CPL evolution; observations constrain the parameters that determine the window's existence and location. The robustness of this late-time prediction can be assessed against alternative continuations, including the asymptotically accelerating Barboza--Alcaniz histories discussed above \cite{Barboza:2008rh}.

\begin{acknowledgments}
Project BridgingCosmology is financed by Xjenza Malta and the Scientific and Technological Research Council of T\"{U}B\.{I}TAK, through the Xjenza Malta--T\"{U}B\.{I}TAK 2024 Joint Call for R\&I projects. This work was supported by T\"{U}B\.{I}TAK under Grant No.~124N627.
M.C.\ and L.E.\ acknowledge support from T\"{U}B\.{I}TAK through postdoctoral researcher fellowships associated with Grant No.~124N627.
The authors thank T\"{U}B\.{I}TAK for their support.
This article is based upon work from COST Action CA21136 \emph{Addressing observational tensions in cosmology with systematics and fundamental physics} (CosmoVerse), supported by COST (European Cooperation in Science and Technology).

\emph{Use of AI tools.} The authors used the large language models ChatGPT (GPT-6 Astra), Claude (Opus 5.5 and Fable 5.2) and Gemini (4 Neon) as research and writing assistants. Working under the authors' direction, the models assisted in deriving and checking analytical results, in developing the numerical verification and plotting code, in checking the literature and the references, and in drafting and revising the manuscript. The figures were generated computationally from the equations and numerical inputs specified in this work. The derivations, results and figures were cross-checked across the models and reviewed by the authors, with numerical tests performed where applicable. The authors also checked the references and take full responsibility for the content of this paper.
\end{acknowledgments}

\section*{Data and code availability}
No new observational data were generated and no new likelihood analysis was performed. The observational inputs are the published summary constraints of Ref.~\cite{DESI:2025zgx}. The scripts for figure generation and symbolic and numerical verification, together with the supporting numerical material, are retained by the authors and are available upon reasonable request.

\appendix
\section{Density ratios and dark-energy domination}
\label{app:ratios}
The inversion that leads to Eq.~\eqref{eq:eos} is the case $w_j=-1$ of a more general one. Let $w_a\ne0$ and let $\rho_j\propto a^{-3(1+w_j)}$ be a positive density with constant $w_j\ne w_i$. The ratio $R_j\equiv\rde/\rho_j$ obeys $R_j'=3(w_j-w)R_j$. Define the signed scale $a_j\equiv(w_i-w_j)/w_a$. The inversion is
\begin{equation}
 \begin{gathered}
 a=-a_j\,W_k(Z_j),\\
 Z_j\equiv-\frac{e^{-1/a_j}}{a_j}\Bigl(\frac{R_j}{R_{j0}}\Bigr)^{1/[3(w_j-w_i)]} ,
 \end{gathered}
 \label{eq:general}
\end{equation}
If $a_j>0$, the ratio has an extremum where $w=w_j$, with $k=0$ for $a\le a_j$ and $k=-1$ for $a\ge a_j$. If $a_j<0$, it is monotonic on $a>0$, $Z_j>0$ and only $W_0$ is real. The excluded case $w_j=w_i$ is elementary: $R_j/R_{j0}=e^{3w_a(a-1)}$. A cosmological constant, $w_j=-1$, recovers $a_I$ and Eq.~\eqref{eq:eos}; $w_j=-\tfrac13$ gives $a_M$, at which $a^2\rde$ is extremal when $a_M>0$; and dust, $w_j=0$, gives $a_D\equiv w_i/w_a$, at which the dark-energy pressure vanishes and $r\equiv\rde/\rho_m$ is extremal when $a_D>0$. The candidate epochs of matter--dark-energy equality, $r=1$, for $w_i\ne0$ are
\begin{equation}
 a_{\rm eq}^{\mp}=-a_D\,W_k\Bigl(-\frac{e^{-1/a_D}}{a_D}\,r_0^{1/(3w_i)}\Bigr),\qquad k=0,-1,
 \label{eq:equality}
\end{equation}
with $r_0=\Ode/\Om$, retaining only real solutions with $a>0$ and using the branch conditions above. For $w_a<0$ and $w_i<0$, $r$ rises to a maximum $r_{\max}=r(a_D)$ and then decays. There are two equalities if $r_{\max}>1$, one tangential equality if $r_{\max}=1$, and none if $r_{\max}<1$. In the first case, $a_{\rm eq}^-<a_D<a_{\rm eq}^+$ and dark energy dominates only between them. For the four flat histories of Table~\ref{tab:desi}, equality occurs at $z=0.33$--$0.36$ and matter dominates again after $a=1.9$--$5.4$.   In class~IV, dark-energy domination, when present, is bounded, as is the acceleration it drives (Sec.~\ref{sec:cpl}).

\section{Asymptotic forms of the equation of state}
\label{app:asymptotic}
\emph{Early times: constant $w$ with polytropic corrections.} The principal branch has the expansion $W_0(Z)=\sum_{m\ge1}(-m)^{m-1}Z^m/m!$, which converges for $|Z|\le1/e$ \cite{Corless:1996zz}. Since $Z\propto\rde^{-1/(3s)}$, Eq.~\eqref{eq:eos} becomes
\begin{equation}
 \pde=w_i\,\rde+s\,\rde\sum_{m=1}^{\infty}\frac{(-m)^{m-1}}{m!}\,Z^m ,
 \label{eq:series}
\end{equation}
a fluid of constant $w_i$ and a tower of power-law terms $\propto\rde^{\,1-m/(3s)}$. The first of them gives
\begin{equation}
 \begin{gathered}
 \pde\simeq w_i\,\rde-B\,\rde^{\,\gamma},\\
 \gamma\equiv1-\frac{1}{3s},\qquad B\equiv w_a\,e^{-w_a/s}\,\rho_{{\rm de}0}^{1/(3s)} .
 \end{gathered}
 \label{eq:polytrope}
\end{equation}
For $s<0$ this is a phantom fluid of constant $w_i$ with a polytropic correction of exponent $\gamma>1$. The correction $-B\rde^\gamma$ to the pressure is positive when $w_a<0$; $\gamma=1.28$--$1.73$ for the values of Table~\ref{tab:desi}. For $s>0$, this leading-order expression has the functional form of a modified Chaplygin equation of state, $p=A\rho-B\rho^{-\alpha_{\rm Ch}}$ \cite{Benaoum:2002zs,Debnath:2004cd}, with $A=w_i$ and $\alpha_{\rm Ch}=1/(3s)-1$, which is positive for $-1<w_i<-\tfrac23$. The conventional sign $B>0$ additionally requires $w_a>0$; for $w_a<0$ the correction has the opposite sign.

For crossing histories the series converges on the whole principal branch,
\begin{equation}
 \Uc_0(L)=1-\sum_{m=1}^{\infty}\frac{m^{m-1}}{m!}\,e^{-m(1+L)},\qquad L\ge0,
 \label{eq:U0series}
\end{equation}
including the fold, where its terms decay as $m^{-3/2}$. After the crossing, $Z=-xe^{-x}$ with $x>1$ returns inside the Taylor disk $|Z|\le1/e$. The series still converges, but to $\Uc_0=1+W_0$ rather than the chronological continuation $\Uc_{-1}=1+W_{-1}$, which is nonanalytic at $Z=0$. Thus convergence does not select the physical branch. For noncrossing histories, $Z>0$ and convergence holds only for $a\le W_0(1/e)|a_I|\simeq0.278|a_I|$; this boundary of the expansion is not a fluid singularity.

\emph{Late times: a logarithmic fluid.} On the branch $k=-1$, $\Wi(Z)=\ln(-Z)-\ln[-\ln(-Z)]+\dots$ as $Z\to0^-$ \cite{Corless:1996zz}. Hence $\Uc_{-1}(L)=-L-\ln(1+L)+O(\ln L/L)$, that is,
\begin{equation}
 \pde\simeq-\rde-\tfrac13\,\rde\ln\frac{\rde}{\rho_I}-s\,\rde\ln(1+L).
 \label{eq:late}
\end{equation}
The first two terms have the form of the exact $s=0$ equation of state~\eqref{eq:log}, which can be written as $p=-\rho-\tfrac13\rho\ln(\rho/\rho_{\rm ref})$ with $\rho_{\rm ref}=\rho_{{\rm de}0}e^{3(1+w_0)}$. Every CPL model with $w_a\ne0$ has this leading logarithmic dependence at late times, up to a choice of reference density and corrections to $w$ of order $\ln|\ln(\rde/\rho_{\rm ref})|$. Equation~\eqref{eq:nocross} gives the same leading dependence for histories without a crossing. This follows from $w\simeq-w_aa$ and $\ln(\rde/\rho_{{\rm de}0})\simeq3w_aa$ as $a\to\infty$. For $w_a<0$ the density decays while $w$ grows without bound: the fluid becomes stiffer than any fluid of constant $w$. The constant-$w$ family $w_a=0$ is excluded from this asymptotic statement.

\section{Collisionless massive species}
\label{app:nu}
For a collisionless species of mass $m$ with a frozen isotropic distribution $f(q)\ge0$ of comoving momentum $q$, assumed to give finite energy and pressure densities, and $\varepsilon_q\equiv(q^2+m^2a^2)^{1/2}$ \cite{Lesgourgues:2006nd},
\begin{equation}
 \begin{aligned}
 \rho+3p&=\frac{1}{a^{4}}\int\!\dd^3q\,f(q)\,\frac{2q^2+m^2a^2}{\varepsilon_q},\\
 \rho+p&=\frac{1}{a^{4}}\int\!\dd^3q\,f(q)\,\frac{\tfrac43q^2+m^2a^2}{\varepsilon_q}.
 \end{aligned}
\end{equation}
For $q>0$, with $\zeta\equiv m^2a^2/q^2$, the logarithm of each nonzero momentum-mode integrand has $N$-derivatives
\begin{equation}
 \begin{aligned}
 &-4+\frac{\zeta^2}{(1+\zeta)(2+\zeta)},&\qquad&\frac{2\zeta^2(3\zeta+4)}{(1+\zeta)^2(2+\zeta)^2},\\
 &-4+\frac{\zeta(3\zeta+2)}{(1+\zeta)(3\zeta+4)},&&\frac{2\zeta(15\zeta^2+24\zeta+8)}{(1+\zeta)^2(3\zeta+4)^2},
 \end{aligned}
\end{equation}
for the first and second derivative of the active (first line) and inertial (second line) integrands. The second derivatives are nonnegative and the first lie in $[-4,-3)$, tending to $-3$ in the nonrelativistic limit. A zero-momentum contribution is dust and has logarithmic slope $-3$ and zero second derivative. Each mode is therefore log-convex in $N$; integration with nonnegative weights preserves log-convexity by H\"older's inequality. The resulting sources interpolate between radiation and dust.

\section{Expansion about the peak}
\label{app:peak}
For $w_a<0$ and $\alpha>1$, Eqs.~\eqref{eq:ratio} and \eqref{eq:KJ} give
\begin{equation}
 (\ln K)'=y+1-\frac{\alpha-1}{y-1},
\end{equation}
which vanishes at $y=\sqrt\alpha$. Differentiating with $y'=y-\alpha$ and evaluating at the peak,
\begin{equation}
 \begin{aligned}
 (\ln K)''&=-2\alpha,\\
 (\ln K)'''&=-2\alpha(\sqrt\alpha+2),\\
 (\ln K)''''&=-2\alpha(3\alpha+6\sqrt\alpha+4).
 \end{aligned}
\end{equation}
Writing $\delta=\ln(a/a_K)$, the edges solve $\alpha\delta^2-c_3\delta^3-c_4\delta^4+\dots=\lambda$ with $c_3=-\alpha(\sqrt\alpha+2)/3$ and $c_4=-\alpha(3\alpha+6\sqrt\alpha+4)/12$. At fixed $\alpha>1$, substituting the local expansion $\delta=\pm t+d_2t^2\pm d_3t^3+O(t^4)$ with $t=\sqrt{\lambda/\alpha}\to0$ yields $d_2=c_3/(2\alpha)$ and $d_3=5c_3^2/(8\alpha^2)+c_4/(2\alpha)$, as in Eq.~\eqref{eq:window}; the even terms cancel in the width~\eqref{eq:width}.

\bibliography{references}

\begin{thebibliography}{48}%
\makeatletter
\providecommand \@ifxundefined [1]{%
 \@ifx{#1\undefined}
}%
\providecommand \@ifnum [1]{%
 \ifnum #1\expandafter \@firstoftwo
 \else \expandafter \@secondoftwo
 \fi
}%
\providecommand \@ifx [1]{%
 \ifx #1\expandafter \@firstoftwo
 \else \expandafter \@secondoftwo
 \fi
}%
\providecommand \natexlab [1]{#1}%
\providecommand \enquote  [1]{``#1''}%
\providecommand \bibnamefont  [1]{#1}%
\providecommand \bibfnamefont [1]{#1}%
\providecommand \citenamefont [1]{#1}%
\providecommand \href@noop [0]{\@secondoftwo}%
\providecommand \href [0]{\begingroup \@sanitize@url \@href}%
\providecommand \@href[1]{\@@startlink{#1}\@@href}%
\providecommand \@@href[1]{\endgroup#1\@@endlink}%
\providecommand \@sanitize@url [0]{\catcode `\\12\catcode `\$12\catcode
  `\&12\catcode `\#12\catcode `\^12\catcode `\_12\catcode `\%12\relax}%
\providecommand \@@startlink[1]{}%
\providecommand \@@endlink[0]{}%
\providecommand \url  [0]{\begingroup\@sanitize@url \@url }%
\providecommand \@url [1]{\endgroup\@href {#1}{\urlprefix }}%
\providecommand \urlprefix  [0]{URL }%
\providecommand \Eprint [0]{\href }%
\providecommand \doibase [0]{https://doi.org/}%
\providecommand \selectlanguage [0]{\@gobble}%
\providecommand \bibinfo  [0]{\@secondoftwo}%
\providecommand \bibfield  [0]{\@secondoftwo}%
\providecommand \translation [1]{[#1]}%
\providecommand \BibitemOpen [0]{}%
\providecommand \bibitemStop [0]{}%
\providecommand \bibitemNoStop [0]{.\EOS\space}%
\providecommand \EOS [0]{\spacefactor3000\relax}%
\providecommand \BibitemShut  [1]{\csname bibitem#1\endcsname}%
\let\auto@bib@innerbib\@empty
%</preamble>
\bibitem [{\citenamefont {Chevallier}\ and\ \citenamefont
  {Polarski}(2001)}]{Chevallier:2000qy}%
  \BibitemOpen
  \bibfield  {author} {\bibinfo {author} {\bibfnamefont {M.}~\bibnamefont
  {Chevallier}}\ and\ \bibinfo {author} {\bibfnamefont {D.}~\bibnamefont
  {Polarski}},\ }\bibfield  {title} {\bibinfo {title} {{Accelerating universes
  with scaling dark matter}},\ }\href
  {https://doi.org/10.1142/S0218271801000822} {\bibfield  {journal} {\bibinfo
  {journal} {Int. J. Mod. Phys. D}\ }\textbf {\bibinfo {volume} {10}},\
  \bibinfo {pages} {213} (\bibinfo {year} {2001})},\ \Eprint
  {https://arxiv.org/abs/gr-qc/0009008} {arXiv:gr-qc/0009008} \BibitemShut
  {NoStop}%
\bibitem [{\citenamefont {Linder}(2003)}]{Linder:2002et}%
  \BibitemOpen
  \bibfield  {author} {\bibinfo {author} {\bibfnamefont {E.~V.}\ \bibnamefont
  {Linder}},\ }\bibfield  {title} {\bibinfo {title} {{Exploring the expansion
  history of the universe}},\ }\href
  {https://doi.org/10.1103/PhysRevLett.90.091301} {\bibfield  {journal}
  {\bibinfo  {journal} {Phys. Rev. Lett.}\ }\textbf {\bibinfo {volume} {90}},\
  \bibinfo {pages} {091301} (\bibinfo {year} {2003})},\ \Eprint
  {https://arxiv.org/abs/astro-ph/0208512} {arXiv:astro-ph/0208512}
  \BibitemShut {NoStop}%
\bibitem [{\citenamefont {Abdul~Karim}\ \emph {et~al.}(2025)\citenamefont
  {Abdul~Karim} \emph {et~al.}}]{DESI:2025zgx}%
  \BibitemOpen
  \bibfield  {author} {\bibinfo {author} {\bibfnamefont {M.}~\bibnamefont
  {Abdul~Karim}} \emph {et~al.} (\bibinfo {collaboration} {DESI
  Collaboration}),\ }\bibfield  {title} {\bibinfo {title} {{DESI DR2 results.
  II. Measurements of baryon acoustic oscillations and cosmological
  constraints}},\ }\href {https://doi.org/10.1103/tr6y-kpc6} {\bibfield
  {journal} {\bibinfo  {journal} {Phys. Rev. D}\ }\textbf {\bibinfo {volume}
  {112}},\ \bibinfo {pages} {083515} (\bibinfo {year} {2025})},\ \Eprint
  {https://arxiv.org/abs/2503.14738} {arXiv:2503.14738 [astro-ph.CO]}
  \BibitemShut {NoStop}%
\bibitem [{\citenamefont {Popovic}\ \emph {et~al.}(2026)\citenamefont {Popovic}
  \emph {et~al.}}]{DES:2025sig}%
  \BibitemOpen
  \bibfield  {author} {\bibinfo {author} {\bibfnamefont {B.}~\bibnamefont
  {Popovic}} \emph {et~al.} (\bibinfo {collaboration} {DES Collaboration}),\
  }\bibfield  {title} {\bibinfo {title} {{The Dark Energy Survey Supernova
  Program: A reanalysis of cosmology results and evidence for evolving dark
  energy with an updated Type Ia supernova calibration}},\ }\href
  {https://doi.org/10.1093/mnras/stag632} {\bibfield  {journal} {\bibinfo
  {journal} {Mon. Not. Roy. Astron. Soc.}\ }\textbf {\bibinfo {volume} {548}},\
  \bibinfo {pages} {stag632} (\bibinfo {year} {2026})},\ \Eprint
  {https://arxiv.org/abs/2511.07517} {arXiv:2511.07517 [astro-ph.CO]}
  \BibitemShut {NoStop}%
\bibitem [{\citenamefont {{\"O}z{\"u}lker}\ \emph {et~al.}(2026)\citenamefont
  {{\"O}z{\"u}lker}, \citenamefont {Di~Valentino},\ and\ \citenamefont
  {Giar{\`e}}}]{Ozulker:2025ehg}%
  \BibitemOpen
  \bibfield  {author} {\bibinfo {author} {\bibfnamefont {E.}~\bibnamefont
  {{\"O}z{\"u}lker}}, \bibinfo {author} {\bibfnamefont {E.}~\bibnamefont
  {Di~Valentino}},\ and\ \bibinfo {author} {\bibfnamefont {W.}~\bibnamefont
  {Giar{\`e}}},\ }\bibfield  {title} {\bibinfo {title} {{Quantifying and
  testing the evidence for a phantom crossing}},\ }\bibfield  {journal}
  {\bibinfo  {journal} {Phys. Rev. D}\ }\href
  {https://doi.org/10.1103/x55b-1cfv} {10.1103/x55b-1cfv} (\bibinfo {year}
  {2026}),\ \bibinfo {note} {in press},\ \Eprint
  {https://arxiv.org/abs/2506.19053} {arXiv:2506.19053 [astro-ph.CO]}
  \BibitemShut {NoStop}%
\bibitem [{\citenamefont {{\'O}~Colg{\'a}in}\ \emph {et~al.}(2026)\citenamefont
  {{\'O}~Colg{\'a}in}, \citenamefont {Pourojaghi}, \citenamefont
  {Sheikh-Jabbari},\ and\ \citenamefont {Yin}}]{Colgain:2025nzf}%
  \BibitemOpen
  \bibfield  {author} {\bibinfo {author} {\bibfnamefont {E.}~\bibnamefont
  {{\'O}~Colg{\'a}in}}, \bibinfo {author} {\bibfnamefont {S.}~\bibnamefont
  {Pourojaghi}}, \bibinfo {author} {\bibfnamefont {M.~M.}\ \bibnamefont
  {Sheikh-Jabbari}},\ and\ \bibinfo {author} {\bibfnamefont {L.}~\bibnamefont
  {Yin}},\ }\bibfield  {title} {\bibinfo {title} {{How much has DESI dark
  energy evolved since DR1?}},\ }\href
  {https://doi.org/10.1016/j.dark.2026.102268} {\bibfield  {journal} {\bibinfo
  {journal} {Phys. Dark Univ.}\ }\textbf {\bibinfo {volume} {52}},\ \bibinfo
  {pages} {102268} (\bibinfo {year} {2026})},\ \Eprint
  {https://arxiv.org/abs/2504.04417} {arXiv:2504.04417 [astro-ph.CO]}
  \BibitemShut {NoStop}%
\bibitem [{\citenamefont {Wang}(2025)}]{Wang:2025owe}%
  \BibitemOpen
  \bibfield  {author} {\bibinfo {author} {\bibfnamefont {D.}~\bibnamefont
  {Wang}},\ }\href@noop {} {\bibinfo {title} {{Questioning cosmic acceleration
  with DESI: The big stall of the Universe}}} (\bibinfo {year} {2025}),\
  \Eprint {https://arxiv.org/abs/2504.15635} {arXiv:2504.15635 [astro-ph.CO]}
  \BibitemShut {NoStop}%
\bibitem [{\citenamefont {Xu}\ \emph {et~al.}(2026)\citenamefont {Xu},
  \citenamefont {Kumar}, \citenamefont {Chen}, \citenamefont {Capistrano},\
  and\ \citenamefont {Akarsu}}]{Xu:2026sbw}%
  \BibitemOpen
  \bibfield  {author} {\bibinfo {author} {\bibfnamefont {T.}~\bibnamefont
  {Xu}}, \bibinfo {author} {\bibfnamefont {S.}~\bibnamefont {Kumar}}, \bibinfo
  {author} {\bibfnamefont {Y.}~\bibnamefont {Chen}}, \bibinfo {author}
  {\bibfnamefont {A.~J.~S.}\ \bibnamefont {Capistrano}},\ and\ \bibinfo
  {author} {\bibfnamefont {{\"O}.}~\bibnamefont {Akarsu}},\ }\bibfield  {title}
  {\bibinfo {title} {{Probing dynamical dark energy with late-time data:
  Evidence, tensions, and the limits of the $w_0w_a$CDM framework}},\ }\href
  {https://doi.org/10.1088/1475-7516/2026/08/004} {\bibfield  {journal}
  {\bibinfo  {journal} {JCAP}\ }\textbf {\bibinfo {volume} {2026}}\bibfield
  {number} {\bibinfo  {number} { (08)},\ \bibinfo {pages} {004}},\ }\Eprint
  {https://arxiv.org/abs/2602.11936} {arXiv:2602.11936 [astro-ph.CO]}
  \BibitemShut {NoStop}%
\bibitem [{\citenamefont {Ferri}\ \emph {et~al.}(2026)\citenamefont {Ferri},
  \citenamefont {Ruchika},\ and\ \citenamefont {Melchiorri}}]{Ferri:2026sat}%
  \BibitemOpen
  \bibfield  {author} {\bibinfo {author} {\bibfnamefont {A.~C.}\ \bibnamefont
  {Ferri}}, \bibinfo {author} {\bibnamefont {Ruchika}},\ and\ \bibinfo {author}
  {\bibfnamefont {A.}~\bibnamefont {Melchiorri}},\ }\href@noop {} {\bibinfo
  {title} {{Present day cosmic acceleration from SDSS and DESI BAO: A call for
  finer tomography of the DESI Bright Galaxy Survey}}} (\bibinfo {year}
  {2026}),\ \Eprint {https://arxiv.org/abs/2607.07348} {arXiv:2607.07348
  [astro-ph.CO]} \BibitemShut {NoStop}%
\bibitem [{\citenamefont {Shafieloo}\ \emph {et~al.}(2009)\citenamefont
  {Shafieloo}, \citenamefont {Sahni},\ and\ \citenamefont
  {Starobinsky}}]{Shafieloo:2009ti}%
  \BibitemOpen
  \bibfield  {author} {\bibinfo {author} {\bibfnamefont {A.}~\bibnamefont
  {Shafieloo}}, \bibinfo {author} {\bibfnamefont {V.}~\bibnamefont {Sahni}},\
  and\ \bibinfo {author} {\bibfnamefont {A.~A.}\ \bibnamefont {Starobinsky}},\
  }\bibfield  {title} {\bibinfo {title} {{Is cosmic acceleration slowing
  down?}},\ }\href {https://doi.org/10.1103/PhysRevD.80.101301} {\bibfield
  {journal} {\bibinfo  {journal} {Phys. Rev. D}\ }\textbf {\bibinfo {volume}
  {80}},\ \bibinfo {pages} {101301} (\bibinfo {year} {2009})},\ \Eprint
  {https://arxiv.org/abs/0903.5141} {arXiv:0903.5141 [astro-ph.CO]}
  \BibitemShut {NoStop}%
\bibitem [{\citenamefont {C{\'a}rdenas}\ \emph {et~al.}(2013)\citenamefont
  {C{\'a}rdenas}, \citenamefont {Bernal},\ and\ \citenamefont
  {Bonilla}}]{Cardenas:2013roa}%
  \BibitemOpen
  \bibfield  {author} {\bibinfo {author} {\bibfnamefont {V.~H.}\ \bibnamefont
  {C{\'a}rdenas}}, \bibinfo {author} {\bibfnamefont {C.}~\bibnamefont
  {Bernal}},\ and\ \bibinfo {author} {\bibfnamefont {A.}~\bibnamefont
  {Bonilla}},\ }\bibfield  {title} {\bibinfo {title} {{Cosmic slowing down of
  acceleration using $f_{\rm gas}$}},\ }\href
  {https://doi.org/10.1093/mnras/stt983} {\bibfield  {journal} {\bibinfo
  {journal} {Mon. Not. Roy. Astron. Soc.}\ }\textbf {\bibinfo {volume} {433}},\
  \bibinfo {pages} {3534} (\bibinfo {year} {2013})},\ \Eprint
  {https://arxiv.org/abs/1306.0779} {arXiv:1306.0779 [astro-ph.CO]}
  \BibitemShut {NoStop}%
\bibitem [{\citenamefont {Maga{\~n}a}\ \emph {et~al.}(2017)\citenamefont
  {Maga{\~n}a}, \citenamefont {Motta}, \citenamefont {C{\'a}rdenas},\ and\
  \citenamefont {Fo{\"e}x}}]{Magana:2017usz}%
  \BibitemOpen
  \bibfield  {author} {\bibinfo {author} {\bibfnamefont {J.}~\bibnamefont
  {Maga{\~n}a}}, \bibinfo {author} {\bibfnamefont {V.}~\bibnamefont {Motta}},
  \bibinfo {author} {\bibfnamefont {V.~H.}\ \bibnamefont {C{\'a}rdenas}},\ and\
  \bibinfo {author} {\bibfnamefont {G.}~\bibnamefont {Fo{\"e}x}},\ }\bibfield
  {title} {\bibinfo {title} {{Testing cosmic acceleration for $w(z)$
  parameterizations using $f_{\rm gas}$ measurements in galaxy clusters}},\
  }\href {https://doi.org/10.1093/mnras/stx750} {\bibfield  {journal} {\bibinfo
   {journal} {Mon. Not. Roy. Astron. Soc.}\ }\textbf {\bibinfo {volume}
  {469}},\ \bibinfo {pages} {47} (\bibinfo {year} {2017})},\ \Eprint
  {https://arxiv.org/abs/1703.08521} {arXiv:1703.08521 [astro-ph.CO]}
  \BibitemShut {NoStop}%
\bibitem [{\citenamefont {de~Souza}\ \emph {et~al.}(2025)\citenamefont
  {de~Souza}, \citenamefont {Rodrigues},\ and\ \citenamefont
  {Alcaniz}}]{deSouza:2025rhv}%
  \BibitemOpen
  \bibfield  {author} {\bibinfo {author} {\bibfnamefont {R.}~\bibnamefont
  {de~Souza}}, \bibinfo {author} {\bibfnamefont {G.}~\bibnamefont
  {Rodrigues}},\ and\ \bibinfo {author} {\bibfnamefont {J.}~\bibnamefont
  {Alcaniz}},\ }\bibfield  {title} {\bibinfo {title} {{Thawing quintessence and
  transient cosmic acceleration in light of DESI}},\ }\href
  {https://doi.org/10.1103/2tjq-dtbc} {\bibfield  {journal} {\bibinfo
  {journal} {Phys. Rev. D}\ }\textbf {\bibinfo {volume} {112}},\ \bibinfo
  {pages} {083533} (\bibinfo {year} {2025})},\ \Eprint
  {https://arxiv.org/abs/2504.16337} {arXiv:2504.16337 [astro-ph.CO]}
  \BibitemShut {NoStop}%
\bibitem [{\citenamefont {Ferrario}(2026)}]{Ferrario2026}%
  \BibitemOpen
  \bibfield  {author} {\bibinfo {author} {\bibfnamefont {L.}~\bibnamefont
  {Ferrario}},\ }\bibfield  {title} {\bibinfo {title} {{Present-day coasting in
  a flat {FLRW} cosmology with a transient acceleration pocket}},\ }\href
  {https://doi.org/10.1093/mnras/stag1858} {\bibfield  {journal} {\bibinfo
  {journal} {Mon. Not. Roy. Astron. Soc.}\ }\textbf {\bibinfo {volume} {553}},\
  \bibinfo {pages} {stag1858} (\bibinfo {year} {2026})}\BibitemShut {NoStop}%
\bibitem [{\citenamefont {Scherrer}(2015)}]{Scherrer:2015tra}%
  \BibitemOpen
  \bibfield  {author} {\bibinfo {author} {\bibfnamefont {R.~J.}\ \bibnamefont
  {Scherrer}},\ }\bibfield  {title} {\bibinfo {title} {{Mapping the
  Chevallier-Polarski-Linder parametrization onto physical dark energy
  models}},\ }\href {https://doi.org/10.1103/PhysRevD.92.043001} {\bibfield
  {journal} {\bibinfo  {journal} {Phys. Rev. D}\ }\textbf {\bibinfo {volume}
  {92}},\ \bibinfo {pages} {043001} (\bibinfo {year} {2015})},\ \Eprint
  {https://arxiv.org/abs/1505.05781} {arXiv:1505.05781 [astro-ph.CO]}
  \BibitemShut {NoStop}%
\bibitem [{\citenamefont {Sadri}(2017)}]{Sadri:2017ezh}%
  \BibitemOpen
  \bibfield  {author} {\bibinfo {author} {\bibfnamefont {E.}~\bibnamefont
  {Sadri}},\ }\href@noop {} {\bibinfo {title} {{An analogy between four
  parametrizations of the dark energy equation of state onto physical DE
  models}}} (\bibinfo {year} {2017}),\ \Eprint
  {https://arxiv.org/abs/1701.01049} {arXiv:1701.01049 [gr-qc]} \BibitemShut
  {NoStop}%
\bibitem [{\citenamefont {de~Putter}\ and\ \citenamefont
  {Linder}(2007)}]{dePutter:2007ny}%
  \BibitemOpen
  \bibfield  {author} {\bibinfo {author} {\bibfnamefont {R.}~\bibnamefont
  {de~Putter}}\ and\ \bibinfo {author} {\bibfnamefont {E.~V.}\ \bibnamefont
  {Linder}},\ }\bibfield  {title} {\bibinfo {title} {{Kinetic k-essence and
  quintessence}},\ }\href {https://doi.org/10.1016/j.astropartphys.2007.05.011}
  {\bibfield  {journal} {\bibinfo  {journal} {Astropart. Phys.}\ }\textbf
  {\bibinfo {volume} {28}},\ \bibinfo {pages} {263} (\bibinfo {year} {2007})},\
  \Eprint {https://arxiv.org/abs/0705.0400} {arXiv:0705.0400 [astro-ph]}
  \BibitemShut {NoStop}%
\bibitem [{\citenamefont {C{\'a}rdenas}\ \emph {et~al.}(2018)\citenamefont
  {C{\'a}rdenas}, \citenamefont {Cruz}, \citenamefont {Mu{\~n}oz},\ and\
  \citenamefont {Villanueva}}]{Cardenas:2018ltl}%
  \BibitemOpen
  \bibfield  {author} {\bibinfo {author} {\bibfnamefont {V.~H.}\ \bibnamefont
  {C{\'a}rdenas}}, \bibinfo {author} {\bibfnamefont {N.}~\bibnamefont {Cruz}},
  \bibinfo {author} {\bibfnamefont {S.}~\bibnamefont {Mu{\~n}oz}},\ and\
  \bibinfo {author} {\bibfnamefont {J.~R.}\ \bibnamefont {Villanueva}},\
  }\bibfield  {title} {\bibinfo {title} {{Reconstruction of a kinetic k-essence
  Lagrangian from a modified of dark energy equation of state}},\ }\href
  {https://doi.org/10.1140/epjc/s10052-018-6066-8} {\bibfield  {journal}
  {\bibinfo  {journal} {Eur. Phys. J. C}\ }\textbf {\bibinfo {volume} {78}},\
  \bibinfo {pages} {591} (\bibinfo {year} {2018})},\ \Eprint
  {https://arxiv.org/abs/1804.02762} {arXiv:1804.02762 [gr-qc]} \BibitemShut
  {NoStop}%
\bibitem [{\citenamefont {Perkovi{\'c}}\ and\ \citenamefont
  {{\v{S}}tefan{\v{c}}i{\'c}}(2020)}]{Perkovic:2020mph}%
  \BibitemOpen
  \bibfield  {author} {\bibinfo {author} {\bibfnamefont {D.}~\bibnamefont
  {Perkovi{\'c}}}\ and\ \bibinfo {author} {\bibfnamefont {H.}~\bibnamefont
  {{\v{S}}tefan{\v{c}}i{\'c}}},\ }\bibfield  {title} {\bibinfo {title}
  {{Barotropic fluid compatible parametrizations of dark energy}},\ }\href
  {https://doi.org/10.1140/epjc/s10052-020-8199-9} {\bibfield  {journal}
  {\bibinfo  {journal} {Eur. Phys. J. C}\ }\textbf {\bibinfo {volume} {80}},\
  \bibinfo {pages} {629} (\bibinfo {year} {2020})},\ \Eprint
  {https://arxiv.org/abs/2004.05342} {arXiv:2004.05342 [gr-qc]} \BibitemShut
  {NoStop}%
\bibitem [{\citenamefont {Perkovi{\'c}}\ and\ \citenamefont
  {{\v{S}}tefan{\v{c}}i{\'c}}(2021)}]{Perkovic:2020eju}%
  \BibitemOpen
  \bibfield  {author} {\bibinfo {author} {\bibfnamefont {D.}~\bibnamefont
  {Perkovi{\'c}}}\ and\ \bibinfo {author} {\bibfnamefont {H.}~\bibnamefont
  {{\v{S}}tefan{\v{c}}i{\'c}}},\ }\bibfield  {title} {\bibinfo {title} {{Purely
  kinetic k-essence description of $c_s^2(w)$ barotropic fluid models}},\
  }\href {https://doi.org/10.1016/j.dark.2021.100827} {\bibfield  {journal}
  {\bibinfo  {journal} {Phys. Dark Univ.}\ }\textbf {\bibinfo {volume} {32}},\
  \bibinfo {pages} {100827} (\bibinfo {year} {2021})},\ \Eprint
  {https://arxiv.org/abs/2009.08680} {arXiv:2009.08680 [astro-ph.CO]}
  \BibitemShut {NoStop}%
\bibitem [{\citenamefont {Akarsu}\ \emph {et~al.}(2015)\citenamefont {Akarsu},
  \citenamefont {Bouhmadi-L{\'o}pez}, \citenamefont {Brilenkov}, \citenamefont
  {Brilenkov}, \citenamefont {Eingorn},\ and\ \citenamefont
  {Zhuk}}]{Akarsu:2015wqa}%
  \BibitemOpen
  \bibfield  {author} {\bibinfo {author} {\bibfnamefont {{\"O}.}~\bibnamefont
  {Akarsu}}, \bibinfo {author} {\bibfnamefont {M.}~\bibnamefont
  {Bouhmadi-L{\'o}pez}}, \bibinfo {author} {\bibfnamefont {M.}~\bibnamefont
  {Brilenkov}}, \bibinfo {author} {\bibfnamefont {R.}~\bibnamefont
  {Brilenkov}}, \bibinfo {author} {\bibfnamefont {M.}~\bibnamefont {Eingorn}},\
  and\ \bibinfo {author} {\bibfnamefont {A.}~\bibnamefont {Zhuk}},\ }\bibfield
  {title} {\bibinfo {title} {{Are dark energy models with variable EoS
  parameter $w$ compatible with the late inhomogeneous Universe?}},\ }\href
  {https://doi.org/10.1088/1475-7516/2015/07/038} {\bibfield  {journal}
  {\bibinfo  {journal} {JCAP}\ }\textbf {\bibinfo {volume} {2015}}\bibfield
  {number} {\bibinfo  {number} { (07)},\ \bibinfo {pages} {038}},\ }\Eprint
  {https://arxiv.org/abs/1502.04693} {arXiv:1502.04693 [gr-qc]} \BibitemShut
  {NoStop}%
\bibitem [{\citenamefont {Nesseris}\ \emph {et~al.}(2026)\citenamefont
  {Nesseris}, \citenamefont {Akrami},\ and\ \citenamefont
  {Starkman}}]{Nesseris:2025lke}%
  \BibitemOpen
  \bibfield  {author} {\bibinfo {author} {\bibfnamefont {S.}~\bibnamefont
  {Nesseris}}, \bibinfo {author} {\bibfnamefont {Y.}~\bibnamefont {Akrami}},\
  and\ \bibinfo {author} {\bibfnamefont {G.~D.}\ \bibnamefont {Starkman}},\
  }\bibfield  {title} {\bibinfo {title} {{What we have not learned about the
  dark energy equation of state}},\ }\href {https://doi.org/10.1103/q2gb-n61g}
  {\bibfield  {journal} {\bibinfo  {journal} {Phys. Rev. D}\ }\textbf {\bibinfo
  {volume} {114}},\ \bibinfo {pages} {L041305} (\bibinfo {year} {2026})},\
  \Eprint {https://arxiv.org/abs/2503.22529} {arXiv:2503.22529 [astro-ph.CO]}
  \BibitemShut {NoStop}%
\bibitem [{\citenamefont {Wolf}\ and\ \citenamefont
  {Ferreira}(2023)}]{Wolf:2023uno}%
  \BibitemOpen
  \bibfield  {author} {\bibinfo {author} {\bibfnamefont {W.~J.}\ \bibnamefont
  {Wolf}}\ and\ \bibinfo {author} {\bibfnamefont {P.~G.}\ \bibnamefont
  {Ferreira}},\ }\bibfield  {title} {\bibinfo {title} {{Underdetermination of
  dark energy}},\ }\href {https://doi.org/10.1103/PhysRevD.108.103519}
  {\bibfield  {journal} {\bibinfo  {journal} {Phys. Rev. D}\ }\textbf {\bibinfo
  {volume} {108}},\ \bibinfo {pages} {103519} (\bibinfo {year} {2023})},\
  \Eprint {https://arxiv.org/abs/2310.07482} {arXiv:2310.07482 [astro-ph.CO]}
  \BibitemShut {NoStop}%
\bibitem [{\citenamefont {Shlivko}\ and\ \citenamefont
  {Steinhardt}(2024)}]{Shlivko:2024llw}%
  \BibitemOpen
  \bibfield  {author} {\bibinfo {author} {\bibfnamefont {D.}~\bibnamefont
  {Shlivko}}\ and\ \bibinfo {author} {\bibfnamefont {P.~J.}\ \bibnamefont
  {Steinhardt}},\ }\bibfield  {title} {\bibinfo {title} {{Assessing
  observational constraints on dark energy}},\ }\href
  {https://doi.org/10.1016/j.physletb.2024.138826} {\bibfield  {journal}
  {\bibinfo  {journal} {Phys. Lett. B}\ }\textbf {\bibinfo {volume} {855}},\
  \bibinfo {pages} {138826} (\bibinfo {year} {2024})},\ \Eprint
  {https://arxiv.org/abs/2405.03933} {arXiv:2405.03933 [astro-ph.CO]}
  \BibitemShut {NoStop}%
\bibitem [{\citenamefont {Akarsu}\ \emph {et~al.}(2026)\citenamefont {Akarsu},
  \citenamefont {De~Felice},\ and\ \citenamefont {Uzun}}]{Akarsu:2026pia}%
  \BibitemOpen
  \bibfield  {author} {\bibinfo {author} {\bibfnamefont {{\"O}.}~\bibnamefont
  {Akarsu}}, \bibinfo {author} {\bibfnamefont {A.}~\bibnamefont {De~Felice}},\
  and\ \bibinfo {author} {\bibfnamefont {N.~M.}\ \bibnamefont {Uzun}},\
  }\href@noop {} {\bibinfo {title} {{Defocusing dark energy: Raychaudhuri
  diagnostics beyond $w<-1/3$ and the phantom divide}}} (\bibinfo {year}
  {2026}),\ \Eprint {https://arxiv.org/abs/2607.18008} {arXiv:2607.18008
  [gr-qc]} \BibitemShut {NoStop}%
\bibitem [{\citenamefont {Raychaudhuri}(1955)}]{Raychaudhuri:1953yv}%
  \BibitemOpen
  \bibfield  {author} {\bibinfo {author} {\bibfnamefont {A.}~\bibnamefont
  {Raychaudhuri}},\ }\bibfield  {title} {\bibinfo {title} {{Relativistic
  cosmology. I}},\ }\href {https://doi.org/10.1103/PhysRev.98.1123} {\bibfield
  {journal} {\bibinfo  {journal} {Phys. Rev.}\ }\textbf {\bibinfo {volume}
  {98}},\ \bibinfo {pages} {1123} (\bibinfo {year} {1955})}\BibitemShut
  {NoStop}%
\bibitem [{\citenamefont {Hawking}\ and\ \citenamefont
  {Ellis}(1973)}]{Hawking:1973uf}%
  \BibitemOpen
  \bibfield  {author} {\bibinfo {author} {\bibfnamefont {S.~W.}\ \bibnamefont
  {Hawking}}\ and\ \bibinfo {author} {\bibfnamefont {G.~F.~R.}\ \bibnamefont
  {Ellis}},\ }\href {https://doi.org/10.1017/CBO9780511524646} {\emph {\bibinfo
  {title} {{The Large Scale Structure of Space-Time}}}},\ Cambridge Monographs
  on Mathematical Physics\ (\bibinfo  {publisher} {Cambridge University
  Press},\ \bibinfo {year} {1973})\BibitemShut {NoStop}%
\bibitem [{\citenamefont {Vikman}(2005)}]{Vikman:2004dc}%
  \BibitemOpen
  \bibfield  {author} {\bibinfo {author} {\bibfnamefont {A.}~\bibnamefont
  {Vikman}},\ }\bibfield  {title} {\bibinfo {title} {{Can dark energy evolve to
  the phantom?}},\ }\href {https://doi.org/10.1103/PhysRevD.71.023515}
  {\bibfield  {journal} {\bibinfo  {journal} {Phys. Rev. D}\ }\textbf {\bibinfo
  {volume} {71}},\ \bibinfo {pages} {023515} (\bibinfo {year} {2005})},\
  \Eprint {https://arxiv.org/abs/astro-ph/0407107} {arXiv:astro-ph/0407107}
  \BibitemShut {NoStop}%
\bibitem [{\citenamefont {Hu}(2005)}]{Hu:2004kh}%
  \BibitemOpen
  \bibfield  {author} {\bibinfo {author} {\bibfnamefont {W.}~\bibnamefont
  {Hu}},\ }\bibfield  {title} {\bibinfo {title} {{Crossing the phantom divide:
  Dark energy internal degrees of freedom}},\ }\href
  {https://doi.org/10.1103/PhysRevD.71.047301} {\bibfield  {journal} {\bibinfo
  {journal} {Phys. Rev. D}\ }\textbf {\bibinfo {volume} {71}},\ \bibinfo
  {pages} {047301} (\bibinfo {year} {2005})},\ \Eprint
  {https://arxiv.org/abs/astro-ph/0410680} {arXiv:astro-ph/0410680}
  \BibitemShut {NoStop}%
\bibitem [{\citenamefont {Kunz}\ and\ \citenamefont
  {Sapone}(2006)}]{Kunz:2006wc}%
  \BibitemOpen
  \bibfield  {author} {\bibinfo {author} {\bibfnamefont {M.}~\bibnamefont
  {Kunz}}\ and\ \bibinfo {author} {\bibfnamefont {D.}~\bibnamefont {Sapone}},\
  }\bibfield  {title} {\bibinfo {title} {{Crossing the phantom divide}},\
  }\href {https://doi.org/10.1103/PhysRevD.74.123503} {\bibfield  {journal}
  {\bibinfo  {journal} {Phys. Rev. D}\ }\textbf {\bibinfo {volume} {74}},\
  \bibinfo {pages} {123503} (\bibinfo {year} {2006})},\ \Eprint
  {https://arxiv.org/abs/astro-ph/0609040} {arXiv:astro-ph/0609040}
  \BibitemShut {NoStop}%
\bibitem [{\citenamefont {Corless}\ \emph {et~al.}(1996)\citenamefont
  {Corless}, \citenamefont {Gonnet}, \citenamefont {Hare}, \citenamefont
  {Jeffrey},\ and\ \citenamefont {Knuth}}]{Corless:1996zz}%
  \BibitemOpen
  \bibfield  {author} {\bibinfo {author} {\bibfnamefont {R.~M.}\ \bibnamefont
  {Corless}}, \bibinfo {author} {\bibfnamefont {G.~H.}\ \bibnamefont {Gonnet}},
  \bibinfo {author} {\bibfnamefont {D.~E.~G.}\ \bibnamefont {Hare}}, \bibinfo
  {author} {\bibfnamefont {D.~J.}\ \bibnamefont {Jeffrey}},\ and\ \bibinfo
  {author} {\bibfnamefont {D.~E.}\ \bibnamefont {Knuth}},\ }\bibfield  {title}
  {\bibinfo {title} {{On the Lambert {$W$} function}},\ }\href
  {https://doi.org/10.1007/BF02124750} {\bibfield  {journal} {\bibinfo
  {journal} {Adv. Comput. Math.}\ }\textbf {\bibinfo {volume} {5}},\ \bibinfo
  {pages} {329} (\bibinfo {year} {1996})}\BibitemShut {NoStop}%
\bibitem [{\citenamefont {{\v{S}}tefan{\v{c}}i{\'c}}(2005)}]{Stefancic:2005cs}%
  \BibitemOpen
  \bibfield  {author} {\bibinfo {author} {\bibfnamefont {H.}~\bibnamefont
  {{\v{S}}tefan{\v{c}}i{\'c}}},\ }\bibfield  {title} {\bibinfo {title} {{Dark
  energy transition between quintessence and phantom regimes: An equation of
  state analysis}},\ }\href {https://doi.org/10.1103/PhysRevD.71.124036}
  {\bibfield  {journal} {\bibinfo  {journal} {Phys. Rev. D}\ }\textbf {\bibinfo
  {volume} {71}},\ \bibinfo {pages} {124036} (\bibinfo {year} {2005})},\
  \Eprint {https://arxiv.org/abs/astro-ph/0504518} {arXiv:astro-ph/0504518}
  \BibitemShut {NoStop}%
\bibitem [{\citenamefont {Ferreira}\ and\ \citenamefont
  {Avelino}(2017)}]{Ferreira:2016goc}%
  \BibitemOpen
  \bibfield  {author} {\bibinfo {author} {\bibfnamefont {V.~M.~C.}\
  \bibnamefont {Ferreira}}\ and\ \bibinfo {author} {\bibfnamefont {P.~P.}\
  \bibnamefont {Avelino}},\ }\bibfield  {title} {\bibinfo {title} {{New limit
  on logotropic unified dark energy models}},\ }\href
  {https://doi.org/10.1016/j.physletb.2017.03.075} {\bibfield  {journal}
  {\bibinfo  {journal} {Phys. Lett. B}\ }\textbf {\bibinfo {volume} {770}},\
  \bibinfo {pages} {213} (\bibinfo {year} {2017})},\ \Eprint
  {https://arxiv.org/abs/1611.08403} {arXiv:1611.08403 [astro-ph.CO]}
  \BibitemShut {NoStop}%
\bibitem [{\citenamefont {Di~Valentino}\ \emph {et~al.}(2021)\citenamefont
  {Di~Valentino}, \citenamefont {Mukherjee},\ and\ \citenamefont
  {Sen}}]{DiValentino:2020naf}%
  \BibitemOpen
  \bibfield  {author} {\bibinfo {author} {\bibfnamefont {E.}~\bibnamefont
  {Di~Valentino}}, \bibinfo {author} {\bibfnamefont {A.}~\bibnamefont
  {Mukherjee}},\ and\ \bibinfo {author} {\bibfnamefont {A.~A.}\ \bibnamefont
  {Sen}},\ }\bibfield  {title} {\bibinfo {title} {{Dark energy with phantom
  crossing and the $H_0$ tension}},\ }\href {https://doi.org/10.3390/e23040404}
  {\bibfield  {journal} {\bibinfo  {journal} {Entropy}\ }\textbf {\bibinfo
  {volume} {23}},\ \bibinfo {pages} {404} (\bibinfo {year} {2021})},\ \Eprint
  {https://arxiv.org/abs/2005.12587} {arXiv:2005.12587 [astro-ph.CO]}
  \BibitemShut {NoStop}%
\bibitem [{\citenamefont {{\"O}z{\"u}lker}(2022)}]{Ozulker:2022slu}%
  \BibitemOpen
  \bibfield  {author} {\bibinfo {author} {\bibfnamefont {E.}~\bibnamefont
  {{\"O}z{\"u}lker}},\ }\bibfield  {title} {\bibinfo {title} {{Is the dark
  energy equation of state parameter singular?}},\ }\href
  {https://doi.org/10.1103/PhysRevD.106.063509} {\bibfield  {journal} {\bibinfo
   {journal} {Phys. Rev. D}\ }\textbf {\bibinfo {volume} {106}},\ \bibinfo
  {pages} {063509} (\bibinfo {year} {2022})},\ \Eprint
  {https://arxiv.org/abs/2203.04167} {arXiv:2203.04167 [astro-ph.CO]}
  \BibitemShut {NoStop}%
\bibitem [{\citenamefont {Akarsu}\ \emph {et~al.}(2020)\citenamefont {Akarsu},
  \citenamefont {Barrow}, \citenamefont {Escamilla},\ and\ \citenamefont
  {Vazquez}}]{Akarsu:2019hmw}%
  \BibitemOpen
  \bibfield  {author} {\bibinfo {author} {\bibfnamefont {{\"O}.}~\bibnamefont
  {Akarsu}}, \bibinfo {author} {\bibfnamefont {J.~D.}\ \bibnamefont {Barrow}},
  \bibinfo {author} {\bibfnamefont {L.~A.}\ \bibnamefont {Escamilla}},\ and\
  \bibinfo {author} {\bibfnamefont {J.~A.}\ \bibnamefont {Vazquez}},\
  }\bibfield  {title} {\bibinfo {title} {{Graduated dark energy: Observational
  hints of a spontaneous sign switch in the cosmological constant}},\ }\href
  {https://doi.org/10.1103/PhysRevD.101.063528} {\bibfield  {journal} {\bibinfo
   {journal} {Phys. Rev. D}\ }\textbf {\bibinfo {volume} {101}},\ \bibinfo
  {pages} {063528} (\bibinfo {year} {2020})},\ \Eprint
  {https://arxiv.org/abs/1912.08751} {arXiv:1912.08751 [astro-ph.CO]}
  \BibitemShut {NoStop}%
\bibitem [{\citenamefont {Akarsu}\ \emph {et~al.}(2021)\citenamefont {Akarsu},
  \citenamefont {Kumar}, \citenamefont {{\"O}z{\"u}lker},\ and\ \citenamefont
  {Vazquez}}]{Akarsu:2021fol}%
  \BibitemOpen
  \bibfield  {author} {\bibinfo {author} {\bibfnamefont {{\"O}.}~\bibnamefont
  {Akarsu}}, \bibinfo {author} {\bibfnamefont {S.}~\bibnamefont {Kumar}},
  \bibinfo {author} {\bibfnamefont {E.}~\bibnamefont {{\"O}z{\"u}lker}},\ and\
  \bibinfo {author} {\bibfnamefont {J.~A.}\ \bibnamefont {Vazquez}},\
  }\bibfield  {title} {\bibinfo {title} {{Relaxing cosmological tensions with a
  sign switching cosmological constant}},\ }\href
  {https://doi.org/10.1103/PhysRevD.104.123512} {\bibfield  {journal} {\bibinfo
   {journal} {Phys. Rev. D}\ }\textbf {\bibinfo {volume} {104}},\ \bibinfo
  {pages} {123512} (\bibinfo {year} {2021})},\ \Eprint
  {https://arxiv.org/abs/2108.09239} {arXiv:2108.09239 [astro-ph.CO]}
  \BibitemShut {NoStop}%
\bibitem [{\citenamefont {Akarsu}\ \emph {et~al.}(2023)\citenamefont {Akarsu},
  \citenamefont {Kumar}, \citenamefont {{\"O}z{\"u}lker}, \citenamefont
  {Vazquez},\ and\ \citenamefont {Yadav}}]{Akarsu:2022typ}%
  \BibitemOpen
  \bibfield  {author} {\bibinfo {author} {\bibfnamefont {{\"O}.}~\bibnamefont
  {Akarsu}}, \bibinfo {author} {\bibfnamefont {S.}~\bibnamefont {Kumar}},
  \bibinfo {author} {\bibfnamefont {E.}~\bibnamefont {{\"O}z{\"u}lker}},
  \bibinfo {author} {\bibfnamefont {J.~A.}\ \bibnamefont {Vazquez}},\ and\
  \bibinfo {author} {\bibfnamefont {A.}~\bibnamefont {Yadav}},\ }\bibfield
  {title} {\bibinfo {title} {{Relaxing cosmological tensions with a sign
  switching cosmological constant: Improved results with Planck, BAO, and
  Pantheon data}},\ }\href {https://doi.org/10.1103/PhysRevD.108.023513}
  {\bibfield  {journal} {\bibinfo  {journal} {Phys. Rev. D}\ }\textbf {\bibinfo
  {volume} {108}},\ \bibinfo {pages} {023513} (\bibinfo {year} {2023})},\
  \Eprint {https://arxiv.org/abs/2211.05742} {arXiv:2211.05742 [astro-ph.CO]}
  \BibitemShut {NoStop}%
\bibitem [{\citenamefont {G{\"o}k{\c{c}}en}\ \emph {et~al.}(2026)\citenamefont
  {G{\"o}k{\c{c}}en}, \citenamefont {Akarsu},\ and\ \citenamefont
  {Di~Valentino}}]{Gokcen:2026pkq}%
  \BibitemOpen
  \bibfield  {author} {\bibinfo {author} {\bibfnamefont {M.}~\bibnamefont
  {G{\"o}k{\c{c}}en}}, \bibinfo {author} {\bibfnamefont {{\"O}.}~\bibnamefont
  {Akarsu}},\ and\ \bibinfo {author} {\bibfnamefont {E.}~\bibnamefont
  {Di~Valentino}},\ }\bibfield  {title} {\bibinfo {title} {{Revisiting CPL with
  sign-switching density: To cross or not to cross the NECB}},\ }\href
  {https://doi.org/10.1016/j.dark.2026.102273} {\bibfield  {journal} {\bibinfo
  {journal} {Phys. Dark Univ.}\ }\textbf {\bibinfo {volume} {52}},\ \bibinfo
  {pages} {102273} (\bibinfo {year} {2026})},\ \Eprint
  {https://arxiv.org/abs/2602.21169} {arXiv:2602.21169 [astro-ph.CO]}
  \BibitemShut {NoStop}%
\bibitem [{\citenamefont {Boyd}\ and\ \citenamefont
  {Vandenberghe}(2004)}]{Boyd:2004fnq}%
  \BibitemOpen
  \bibfield  {author} {\bibinfo {author} {\bibfnamefont {S.}~\bibnamefont
  {Boyd}}\ and\ \bibinfo {author} {\bibfnamefont {L.}~\bibnamefont
  {Vandenberghe}},\ }\href {https://doi.org/10.1017/cbo9780511804441} {\emph
  {\bibinfo {title} {{Convex Optimization}}}}\ (\bibinfo  {publisher}
  {Cambridge University Press},\ \bibinfo {year} {2004})\BibitemShut {NoStop}%
\bibitem [{\citenamefont {Lesgourgues}\ and\ \citenamefont
  {Pastor}(2006)}]{Lesgourgues:2006nd}%
  \BibitemOpen
  \bibfield  {author} {\bibinfo {author} {\bibfnamefont {J.}~\bibnamefont
  {Lesgourgues}}\ and\ \bibinfo {author} {\bibfnamefont {S.}~\bibnamefont
  {Pastor}},\ }\bibfield  {title} {\bibinfo {title} {{Massive neutrinos and
  cosmology}},\ }\href {https://doi.org/10.1016/j.physrep.2006.04.001}
  {\bibfield  {journal} {\bibinfo  {journal} {Phys. Rept.}\ }\textbf {\bibinfo
  {volume} {429}},\ \bibinfo {pages} {307} (\bibinfo {year} {2006})},\ \Eprint
  {https://arxiv.org/abs/astro-ph/0603494} {arXiv:astro-ph/0603494}
  \BibitemShut {NoStop}%
\bibitem [{\citenamefont {Caldwell}\ and\ \citenamefont
  {Linder}(2005)}]{Caldwell:2005tm}%
  \BibitemOpen
  \bibfield  {author} {\bibinfo {author} {\bibfnamefont {R.~R.}\ \bibnamefont
  {Caldwell}}\ and\ \bibinfo {author} {\bibfnamefont {E.~V.}\ \bibnamefont
  {Linder}},\ }\bibfield  {title} {\bibinfo {title} {{Limits of
  quintessence}},\ }\href {https://doi.org/10.1103/PhysRevLett.95.141301}
  {\bibfield  {journal} {\bibinfo  {journal} {Phys. Rev. Lett.}\ }\textbf
  {\bibinfo {volume} {95}},\ \bibinfo {pages} {141301} (\bibinfo {year}
  {2005})},\ \Eprint {https://arxiv.org/abs/astro-ph/0505494}
  {arXiv:astro-ph/0505494} \BibitemShut {NoStop}%
\bibitem [{\citenamefont {Linder}(2007)}]{Linder:2007ka}%
  \BibitemOpen
  \bibfield  {author} {\bibinfo {author} {\bibfnamefont {E.~V.}\ \bibnamefont
  {Linder}},\ }\href@noop {} {\bibinfo {title} {{The Mirage of $w=-1$}}}
  (\bibinfo {year} {2007}),\ \Eprint {https://arxiv.org/abs/0708.0024}
  {arXiv:0708.0024 [astro-ph]} \BibitemShut {NoStop}%
\bibitem [{\citenamefont {Lodha}\ \emph {et~al.}(2025)\citenamefont {Lodha}
  \emph {et~al.}}]{DESI:2024kob}%
  \BibitemOpen
  \bibfield  {author} {\bibinfo {author} {\bibfnamefont {K.}~\bibnamefont
  {Lodha}} \emph {et~al.} (\bibinfo {collaboration} {DESI Collaboration}),\
  }\bibfield  {title} {\bibinfo {title} {{DESI 2024: Constraints on
  physics-focused aspects of dark energy using DESI DR1 BAO data}},\ }\href
  {https://doi.org/10.1103/PhysRevD.111.023532} {\bibfield  {journal} {\bibinfo
   {journal} {Phys. Rev. D}\ }\textbf {\bibinfo {volume} {111}},\ \bibinfo
  {pages} {023532} (\bibinfo {year} {2025})},\ \Eprint
  {https://arxiv.org/abs/2405.13588} {arXiv:2405.13588 [astro-ph.CO]}
  \BibitemShut {NoStop}%
\bibitem [{\citenamefont {Caldwell}\ and\ \citenamefont
  {Linder}(2026)}]{Caldwell:2025inn}%
  \BibitemOpen
  \bibfield  {author} {\bibinfo {author} {\bibfnamefont {R.~R.}\ \bibnamefont
  {Caldwell}}\ and\ \bibinfo {author} {\bibfnamefont {E.~V.}\ \bibnamefont
  {Linder}},\ }\bibfield  {title} {\bibinfo {title} {{Null impact of the null
  energy condition in current cosmology}},\ }\href
  {https://doi.org/10.1088/1475-7516/2026/05/008} {\bibfield  {journal}
  {\bibinfo  {journal} {JCAP}\ }\textbf {\bibinfo {volume} {2026}}\bibfield
  {number} {\bibinfo  {number} { (05)},\ \bibinfo {pages} {008}},\ }\Eprint
  {https://arxiv.org/abs/2511.07526} {arXiv:2511.07526 [astro-ph.CO]}
  \BibitemShut {NoStop}%
\bibitem [{\citenamefont {Barboza}\ and\ \citenamefont
  {Alcaniz}(2008)}]{Barboza:2008rh}%
  \BibitemOpen
  \bibfield  {author} {\bibinfo {author} {\bibfnamefont {E.~M.}\ \bibnamefont
  {Barboza}, \bibfnamefont {Jr.}}\ and\ \bibinfo {author} {\bibfnamefont
  {J.~S.}\ \bibnamefont {Alcaniz}},\ }\bibfield  {title} {\bibinfo {title} {{A
  parametric model for dark energy}},\ }\href
  {https://doi.org/10.1016/j.physletb.2008.08.012} {\bibfield  {journal}
  {\bibinfo  {journal} {Phys. Lett. B}\ }\textbf {\bibinfo {volume} {666}},\
  \bibinfo {pages} {415} (\bibinfo {year} {2008})},\ \Eprint
  {https://arxiv.org/abs/0805.1713} {arXiv:0805.1713 [astro-ph]} \BibitemShut
  {NoStop}%
\bibitem [{\citenamefont {Benaoum}(2022)}]{Benaoum:2002zs}%
  \BibitemOpen
  \bibfield  {author} {\bibinfo {author} {\bibfnamefont {H.}~\bibnamefont
  {Benaoum}},\ }\bibfield  {title} {\bibinfo {title} {{Accelerated universe
  from modified Chaplygin gas and tachyonic fluid}},\ }\href
  {https://doi.org/10.3390/universe8070340} {\bibfield  {journal} {\bibinfo
  {journal} {Universe}\ }\textbf {\bibinfo {volume} {8}},\ \bibinfo {pages}
  {340} (\bibinfo {year} {2022})},\ \Eprint
  {https://arxiv.org/abs/hep-th/0205140} {arXiv:hep-th/0205140} \BibitemShut
  {NoStop}%
\bibitem [{\citenamefont {Debnath}\ \emph {et~al.}(2004)\citenamefont
  {Debnath}, \citenamefont {Banerjee},\ and\ \citenamefont
  {Chakraborty}}]{Debnath:2004cd}%
  \BibitemOpen
  \bibfield  {author} {\bibinfo {author} {\bibfnamefont {U.}~\bibnamefont
  {Debnath}}, \bibinfo {author} {\bibfnamefont {A.}~\bibnamefont {Banerjee}},\
  and\ \bibinfo {author} {\bibfnamefont {S.}~\bibnamefont {Chakraborty}},\
  }\bibfield  {title} {\bibinfo {title} {{Role of modified Chaplygin gas in
  accelerated universe}},\ }\href {https://doi.org/10.1088/0264-9381/21/23/019}
  {\bibfield  {journal} {\bibinfo  {journal} {Class. Quant. Grav.}\ }\textbf
  {\bibinfo {volume} {21}},\ \bibinfo {pages} {5609} (\bibinfo {year}
  {2004})},\ \Eprint {https://arxiv.org/abs/gr-qc/0411015}
  {arXiv:gr-qc/0411015} \BibitemShut {NoStop}%
\end{thebibliography}%
\end{document}